\documentclass[twocolumn]{svjour3} 

\smartqed 
\PassOptionsToPackage{table}{xcolor}
\usepackage{tikz}
\usetikzlibrary{shapes}
\usepackage{graphicx}
\usepackage{lmodern}
\usepackage{booktabs}
\usepackage{algorithm,algpseudocode}
\usepackage{float}
\usepackage{amssymb}
\usepackage{amsmath}
\usepackage{bm}
\usepackage{hyperref}
\usepackage{natbib}
\usepackage{lineno}
\usepackage{soul}
\usepackage{placeins}
\usepackage{subfigure}

\usepackage{tabularx}
\usepackage[range-phrase = --,range-units = single, product-units = power, per-mode = symbol]{siunitx}
\DeclareSIUnit\pixel{px}
\DeclareSIUnit\voxel{vox}

\usepackage{multirow}

\usepackage{comment}

\begin{document}
\definecolor{color1}{rgb}{0.267004, 0.004874, 0.329415}
\definecolor{color2}{rgb}{0.229739, 0.322361, 0.545706}
\definecolor{color3}{rgb}{0.127568, 0.566949, 0.550556}
\definecolor{color4}{rgb}{0.369214, 0.788888, 0.382914}
\definecolor{color5}{rgb}{0.993248, 0.906157, 0.143936}

\newcommand{\markerone}{\raisebox{0.0pt}{\tikz{\node[draw,scale=0.6,circle,fill=color1, line width=0.02cm](){};}}}
\newcommand{\markertwo}{\raisebox{0pt}{\tikz{\node[draw,scale=0.6,regular polygon, regular polygon sides=4,fill=color2, line width=0.02cm](){};}}}
\newcommand{\markerthree}{\raisebox{-1.0pt}{\tikz{\node[draw,scale=0.6,regular polygon, regular polygon sides=4,fill=color3,rotate=45, line width=0.02cm](){};}}}
\newcommand{\markerfour}{\raisebox{-0.5pt}{\tikz{\node[draw,scale=0.45,regular polygon, regular polygon sides=3,fill=color4,rotate=180, line width=0.02cm](){};}}}
\newcommand{\markerfive}{\raisebox{-0.7pt}{\tikz{\node[draw,scale=0.35,star,star point ratio=2.5,fill=color5, line width=0.02cm](){};}}}

\sloppy
\title{A meshless and binless approach to compute statistics in 3D Ensemble PTV}

\author{Manuel Ratz and  Miguel A. Mendez}

\institute{ Manuel Ratz \at von Karman Institute for Fluid Dynamics, Belgium
              \email{manuel.ratz@vki.ac.be}
              \and
              Miguel Alfonso Mendez \at von Karman Institute for Fluid Dynamics, Belgium              \email{miguel.alfonso.mendez@vki.ac.be}
}

\date{\today}

\maketitle
\abstract{
We propose a method to obtain super\-resolution of turbulent statistics for three-\-dimensional ensemble particle tracking velocimetry (EPTV). The method is ``meshless'' because it does not require the definition of a grid for computing derivatives, and it is ``binless'' because it does not require the definition of bins to compute local statistics. The method combines the constrained radial basis function (RBF) formalism introduced Sperotto et al. (Meas Sci Technol, 33:094005, 2022) with an ensemble trick for the RBF regression of flow statistics. The computational cost for the RBF regression is alleviated using the partition of unity method (PUM).
Three test cases are considered: (1) a 1D illustrative problem on a Gaussian process, (2) a 3D synthetic test case reproducing a 3D jet-like flow, and (3) an experimental dataset collected for an underwater jet flow at $\text{Re} = 6750$ using a four-camera 3D PTV system.
For each test case, the method performances are compared to traditional binning approaches such as Gaussian weighting (Agüí and Jiménez, JFM, 185:447-468, 1987), local polynomial fitting (Agüera et al, Meas Sci Technol, 27:124011, 2016), as well as binned versions of RBFs.}

\vspace{-5mm}
\section{Introduction}\label{sec:introduction}

Much research has focused on developing image-based three-dimensional and three-component velocity measurements (3D3C \cite{Scarano2013}) in the last two decades. The first popular 3D3C technique is the tomographic particle image velocimetry (PIV) introduced by \cite{Elsinga2006}. This extends the planar cross-correlation-based PIV to a three-dimensional setting, where interrogation windows are replaced by interrogation volumes. The main limitation is the computational cost, which scales poorly when moving from 2D to 3D, and the unavoidable spatial filtering produced by a correlation-based evaluation. Recently, 3D Particle Tracking Velocimetry (PTV) has emerged as a promising alternative, offering better computational performances and much higher spatial resolution \citep{Kaehler2012a, Kaehler2012b, Kaehler2016, Schroeder2023}.

A key enabler to the success of 3D PTV has been the development of advanced tracking algorithms such as Shake-the-Box \citep{Schanz2016} or its open-source variant \citep{Tan2020}. These, together with advancements in the particle reconstruction process \citep{Wieneke2013, Schanz2013, Jahn2021}, allow processing images with a particle seeding concentration up to 0.125\,particles per pixel (ppp),  well above the limits of 0.005 ppp of early tracking methods \citep{Maas1993}. Nevertheless, PTV processing produces randomly scattered data. This poses many challenges to post-processing, from the simple computation of gradients (e.g. to compute vorticity) and flow statistics to more advanced pressure integration. Although post-processing methods based on unstructured grids have been proposed (see \cite{Neeteson2015,Neeteson2016}), the most common approach is to interpolate the scattered data onto a uniform grid that allows using traditional post-processing approaches (e.g. finite differences for derivatives, ensemble statistics, modal decompositions etc.).

When interpolation onto Cartesian grid aims at treating instantaneous fields, for example for derivative computations and/or pressure reconstruction, the most popular approaches are Vic+ and Vic\# \citep{Schneiders2016, Scarano2022, Jeon2022}, constrained cost minimization \citep{Agarwal2021}, the FlowFit algorithm \citep{Schanz2016,Gesemann2016}, or Meshless Track Assimilation \cite{Sperotto2024b}. These methods require time-resolved data and introduce some physics-based penalty or constraint to make the interpolation more robust. Examples are the divergence-free condition on the velocity fields or the curl-free condition for acceleration and pressure fields.
 
When interpolation onto Cartesian grids aims at computing flow statistics, such as mean fields or Reynolds stresses, the most popular approaches are based on the concept of binning and ensemble PTV (EPTV, \citealt{Discetti2018}). This method involves dividing the measurement domain into bins, within which local statistics are computed by treating all samples in a bin as part of a local distribution \citep{Kaehler2012a}. If sufficiently dense clouds of points are available, these methods can significantly outperform cross-correlation-based approaches in computing Reynolds stresses \citep{Proebsting2013, Atkinson2014, Schroeder2018}.

EPTV methods vary in how local statistics, particularly second or higher-order moments, are computed. A traditional approach, often called "top-hat," assigns equal weight to all samples within a bin. In contrast, the more advanced Gaussian weighting method by \cite{Aguei1987} assigns greater weight to samples closer to the bin center. \citet{Godbersen2020} demonstrated that integrating a fit of individual particle tracks significantly improves convergence. However, this approach requires particle tracks over multiple time steps, obtained either from time-resolved measurements or multi-pulse data \citep{Novara2016}. \cite{Agueera2016} demonstrated that the top-hat approach suffers from unresolved velocity gradients, while Gaussian weighting results in slower statistical convergence. These issues are exacerbated in three-dimensional EPTV, where achieving statistical convergence may require an impractically large number of samples. To address these limitations, \cite{Agueera2016} proposed using local polynomial fits within each bin to regress the mean flow and then compute higher-order statistics on the mean-subtracted fields. This method combines \emph{spatial} averaging with \emph{ensemble} averaging, allowing for a larger bin size (which benefits statistical convergence) without compromising the resolution of gradients in the mean flow. However, the mean flow is only locally smooth, does not account for physical priors and provides statistics only at the bin's centers.

In this work, we aim to extend the concept of blending and integrate it with the mesh-less framework proposed by \cite{Sperotto2022a, Ratz2022a}, recently released in an open-source toolbox called SPICY (Super-resolution and Pressure from Image veloCimetrY, \citet{Sperotto2024}). The mesh-less approach is a new paradigm in PTV data post-processing, where the interpolation step is entirely removed, and \emph{all} post-processing operations (such as computations of derivatives, correlations, or pressure fields) are performed analytically. In the approach proposed by \cite{Sperotto2022a}, the analytic representation is built using physics-constrained radial basis functions (RBFs). The goal of operating on analytically (symbolically differentiable) fields bridges assimilation methods in velocimetry with machine learning-based super-resolution techniques, including deep learning \citep{Park2020}, Physics-Informed Neural Networks (PINNs, \citealt{Rao2020}), and generative adversarial networks \citep{Gueemes2022}. The primary advantage of the RBF formulation is its linearity with respect to the training parameters, allowing for efficient training and implementation of hard constraints.

The approach proposed in this work employs the constrained RBF framework for spatial averaging, similar to the local polynomial regression by \cite{Agueera2016}. However, we use an ensemble trick to avoid the need for defining bins, resulting in an analytic expression for the statistical quantities that is both grid-free and bin-free. The general formulation is presented in Section \ref{sec:2}. Section \ref{sec:3} outlines the main numerical recipes to implement the RBF constraints while significantly reducing computational costs compared to the original implementation in \cite{Sperotto2022a}. This is achieved using a simplified version of the well-known partition of unity method (PUM, \cite{Melenk1996}) for RBF regression (see also \citet{Larsson2017,Cavoretto2019,Cavoretto2020}). The PUM significantly reduces memory and computational demands by splitting the domain into patches, performing RBF regression in each patch, and then merging the solutions into a single regression. The RBF-PUM was recently applied for super-resolution of Shake-the-Box measurements \citep{Li2021} and mean flow fields in microfluidics \citep{Ratz2022b}, though without penalties or constraints. Section \ref{sec:5} presents the selected test cases for evaluating the algorithm's performance, while Section \ref{sec:4} reviews the algorithms used for benchmarking. Results are presented in Section \ref{sec:6}, and conclusions and perspectives are discussed in Section \ref{sec:7}.

\section{Bin-Free Statistics}\label{sec:2}

We briefly review the fundamentals of Radial Basis Function (RBF) regression in Subsection \ref{sec:2p1}. Subsection \ref{sec:2p2} introduces the ensemble trick to circumvent the need for binning. 

\vspace{-3mm}
\subsection{Fundamentals of RBF regression and notation}\label{sec:2p1}
The RBF regression consists of approximating a function as a linear combination of radial basis functions. In this work, we are interested in approximating the components of 3D velocity fields and consider only isotropic Gaussian RBFs:  
\begin{equation}
    \varphi_k(\mathbf{x}\vert\mathbf{x}_{c,k}, c_k) = \text{exp}\left( -c_k^2 \vert\vert \mathbf{x} - \mathbf{x}_{c,k} \vert\vert^2 \right),
    \label{eq:gaussian_rbfs}
\end{equation} where $\mathbf{x}=(x,y,z) \in \mathbb{R}^3$ is the coordinate where the basis is evaluated, $\mathbf{x}_{c,k}=(x_{c,k},y_{c,k},z_{c,k}) \in \mathbb{R}^3$ and $c_k$ are respectively the $k$-th collocation point and the shape parameter of the basis, and $||\bullet||$ denote the $l_2$ norm of a vector.

At any given point $\mathbf{x}$, the velocity field has three entries $\mathbf{u}(\mathbf{x})=(u(\mathbf{x}),v(\mathbf{x}),w(\mathbf{x})) \in \mathbb{R}^3$. The RBF regression using $n_b$ RBFs can be written as:
\begin{equation}
\label{RBF}
\mathbf{u}(\mathbf{x}) = \begin{pmatrix}
    u \\
    v \\
    w
\end{pmatrix} \approx \sum_{k=1}^{n_b} \begin{pmatrix}
    w_{u,k}\, \varphi_k (\mathbf{x} \vert \mathbf{x}_{c,k}, c_k) \\
    w_{v,k}\, \varphi_k (\mathbf{x} \vert \mathbf{x}_{c,k}, c_k) \\
    w_{w,k}\, \varphi_k (\mathbf{x} \vert \mathbf{x}_{c,k}, c_k)
\end{pmatrix}\,,
\end{equation} where $w_{u,k}, w_{v,k}, w_{w,k} \in \mathbb{R}^{n_b}$ are the weights associated to each basis. The function approximation \eqref{RBF} can conveniently be evaluated on an arbitrary set of points $\bm{X} = (\bm{x}, \bm{y}, \bm{z})$, with $\bm{x},\bm{y},\bm{z}\in\mathbb{R}^{n_p}$ the vectors collecting the coordinates in each point, using the basis matrix $\bm{\Phi}_b(\bm{X})$: this collects the value of each RBF on a set of points $\bm{X}$:
\begin{equation}
\label{eq:basis_matrix}
\bm{\Phi}_b(\bm{X}) = \begin{pmatrix} \vdots & \dots& \vdots \\\varphi(\bm{X};\mathbf{x}_{c,1},c_1) & \dots & \varphi(\bm{X};\mathbf{x}_{c,n_b},c_{n_b})\\\vdots & \dots& \vdots\\ \end{pmatrix} \,,
\end{equation} 

This matrix allows us to express approximation \eqref{RBF} in a compact notation:
\begin{align}
\label{eq:block_struct}
    \begin{split}
        \bm{U}(\bm{X}) &= \begin{pmatrix}
            \bm{u}(\bm{X}) \\
            \bm{v}(\bm{X}) \\
            \bm{w}(\bm{X})
        \end{pmatrix}\\ &\approx \begin{pmatrix}
            \bm{\Phi}_b(\bm{X}) & \mathbf{0}  & \mathbf{0} \\
            \mathbf{0} & \bm{\Phi}_b(\bm{X}) & \mathbf{0} \\
            \mathbf{0} & \mathbf{0} & \bm{\Phi}_b(\bm{X}) 
        \end{pmatrix} \begin{pmatrix}
            \bm{w}_u \\
            \bm{w}_v \\
            \bm{w}_w \\
        \end{pmatrix}\,.
    \end{split}
\end{align} 

This block structure is useful when constraints are introduced later on. To ease the notation, we abbreviate \eqref{eq:block_struct} to $\bm{U}(\bm{X}) \approx \bm{\Phi}(\bm{X}) \bm{W}$. Here, it is understood that $\bm{U} \in \mathbb{R}^{3n_p}$ and $\bm{W} \in\mathbb{R}^{3n_b}$ are the vertically concatenated velocity field and weights, respectively. 

We assume that training data (e.g. PTV measurements) are available on a set of $\bm{X}_*=(\bm{x}_*,\bm{y}_*,\bm{z}_*) \in \mathbb{R}^{3\times n_*}$ points and denote these samples as $\bm{U}(\bm{X}_*)=\bm{U}_*=(\bm{u}_*;\bm{v}_*;\bm{w}_*)\in\mathbb{R}^{3n_*}$ where `;' denotes vertical concatenation. With the basis matrix $\bm{\Phi}(\bm{X}_*)=\bm{\Phi}_*$, the weights minimizing the $l_2$ norm of the training error are (see for example \cite{Hastie2009, Bishop, Deisenroth}):
\begin{equation}
\label{RBF_SOL}
\bm{W}=\bigl(\bm{\Phi}^T_*\bm{\Phi}_*+\alpha\bm{I}\bigr)^{-1} \bm{\Phi}^T_*\bm{U}_*\,,
\end{equation} where $\alpha\in\mathbb{R}$ is a regularization parameter and $\bm{I}$ is the identity matrix. The regularization parameter $\alpha$ ensures that the inversion is possible. Once the weights are computed, the velocity field and its derivatives are available on an arbitrary grid since \eqref{RBF} gives an analytical expression \citep{Sperotto2022a}.

\vspace{-3mm}
\subsection{From ensembles of RBFs to RBF of the ensemble}\label{sec:2p2}

Let us consider a statistically stationary and ergodic velocity field $\mathbf{u}(\mathbf{x})$. The sample at any location $\mathbf{x}$ depends on the joint probability density function (pdf) $f_u(\mathbf{x},\bm{u})$, so that we can define the mean field from the expectation operator: 
\begin{equation}
\label{MEAN}
\langle\mathbf{u}\rangle(\mathbf{x})=\mathbb{E}\{\mathbf{u}(\mathbf{x})\}=\int^{\infty}_{-\infty} \mathbf{u}(\mathbf{x}) f_u(\mathbf{x},\mathbf{u}) \text{d}\mathbf{u}\,.
\end{equation}

The challenge in estimating the mean field in \eqref{MEAN} from a set of PTV measurements of the velocity field is that each sample (snapshot) is available on a different set of points. We denote as $\bm{X}^{(j)}$ the set of $n^{(j)}_p$ points at which the data are available in the $j$-th sample of the field (i.e. PTV measurements) and as $\bm{U}^{(j)}=\bm{U}\bigl(\bm{X}^{(j)}\bigr)$ the associated velocity measurements.

The usual binning-based approach to compute statistics maps the sets of points $\bm{X}^{(j)}$ onto a fixed grid of bins so that all points within the bins can be used to build \emph{local} statistical estimates. Then, attributing all points within the $i$-th bin to a specific position $\bm{x}_i$ allows to remove the spatial dependency of the joint pdf and to move from the expectation operator in \eqref{MEAN} to its discrete (sample-based) counterpart. Therefore, at each of the bin locations $\bm{x}_i$ one has:
\begin{equation}
\langle\mathbf{u}\rangle(\mathbf{x}_i)\approx  \frac{1}{n_{p,i}} \sum^{n_{p,i}}_{i=1} \bm{U}^{(j)}(\mathbf{x}_i)\,,
\end{equation} where $\bm{U}^{(j)}(\mathbf{x}_i)$ denotes the mapping of the PTV sample $\bm{U}^{(j)}$ onto the $i$-th bin, and $n_{p,i}$ denotes the number of measurement points available within the bin.

In this work, we propose an alternative path.
Introducing the RBF regression \eqref{eq:block_struct} into \eqref{MEAN} and noticing that the Jacobian is $\text{d} \mathbf{u}/\text{d}\bm{W}=\bm{\Phi}(\mathbf{x})$ we have: 
\begin{equation}
\label{eq:stats_from_rbf}
\begin{split}
\langle\mathbf{u}\rangle(\mathbf{x})&=\int^{\infty}_{\infty} \bm{\Phi}(\mathbf{x})\bm{W} f_u (\mathbf{x}, \bm{\Phi}(\mathbf{x})) \text{d}\mathbf{u}\\
&=\bm{\Phi}(\mathbf{x})\int^{\infty}_{\infty} \bm{W} f_w (\bm{W}) \text{d}\bm{W}=\,\bm{\Phi}(\mathbf{x}) \langle\bm{W}\rangle\,,
\end{split}
\end{equation} with $f_w(\bm{W})=f_u(\mathbf{x},\mathbf{u})\bm{\Phi}(\mathbf{x})$.
The expectation of the weights can be estimated from data more easily than the expectation of the velocity field, because the distribution $f_w(\bm{W})$ does not depend, at least in principle, on the positioning of the data so long as the regression is successful. This implies that the same set of RBFs is used for the regression of all snapshots and that each of these is sufficiently dense to ensure good training. 

Assuming that one collects $n_t$ velocity fields and denoting as $\bm{W}^{(j)}$ the weights of the RBF regression of each of the $j=[1,\dots n_t]$ snapshots, one has:
\begin{equation}
\label{w_bar}
\langle\bm{W}\rangle \approx\bm{W}_A = \frac{1}{n_t} \sum^{n_t}_{j=1} \bm{W}^{(j)} \,,
\end{equation} where $\bm{W}^{(j)}$ is the weight obtained when regressing the $j$-th snapshot. Introducing \eqref{RBF_SOL} is particularly revealing:
\begin{equation}
    \label{light}
\langle\bm{W}\rangle\approx \frac{1}{n_t} \sum^{n_t}_{j=1}  \bigl(\bm{\Phi}^T_{(j)}\bm{\Phi}_{(j)}+\alpha\bm{I}\bigr)^{-1} \bm{\Phi}^T_{(j)}\bm{u}^{(j)}\,,
\end{equation} where $\bm{\Phi}_{(j)}=\bm{\Phi}\bigl(\bm{X}^{(j)}\bigr)$. All operations in \eqref{light} are linear, and some of these can be replaced by operations on the full ensemble of data, which we define as:
\begin{equation}
\bm{X}_E=\bigcup_{j \in 1\dots n_t} \bm{X}_{(j)} \quad \quad \bm{U}_E= \bigcup_{j \in 1\dots n_t} \bm{U}_{(j)}\,.
\end{equation}  

The ensemble has useful properties. Defining as $\bm{\Phi}_E=\bm{\Phi}(\bm{X}_E)$ and expanding the summations in the projections $\bm{\Phi}^T_{(j)}\bm{U}^{(j)}$ and in the correlations $\bm{\Phi}^T_{(j)}\bm{\Phi}_{(j)}$ from \eqref{light}, one has:
\begin{subequations}
\begin{equation}
\label{corr_f}
\sum^{n_t}_{j=1}\bm{\Phi}^T_{(j)}\bm{\Phi}_{(j)}= \bm{\Phi}^T_E\bm{\Phi}_E \,\in \mathbb{R}^{n_b\times n_b}\,,
\end{equation}    
\begin{equation}
\label{corr_f2}
\sum^{n_t}_{j=1}\bm{\Phi}^T_{(j)}\bm{U}^{(j)} = \bm{\Phi}^T_E \bm{U}_E\, \,\in \mathbb{R}^{n_b\times n_b}\,.
\end{equation}
\end{subequations}

The goal of the proposed approach is to replace the average of the RBF regression in each snapshot, as requested in \eqref{light}, with the RBF regression of the ensemble set. This allows replacing $n_t$ regressions of size $n^{(j)}_p$ with one single regression of size $n_{pE}$. Without aiming for a formal proof, we note that the covariance matrices $\bm{\Phi}^T_{(j)}\bm{\Phi}_{(j)}$ collect the inner products between the bases sampled on the points $\bm{X}^{(j)}$:
\begin{equation}
{\bm{\Phi}^T_{(j)}\bm{\Phi}_{(j)}}[m,n]=\sum^{n^{(j)}_p}_{j=1} \varphi_m\bigl(\bm{X}^{(j)}\bigr)\varphi_n\bigl(\bm{X}^{(j)}\bigr)\,,
\end{equation} and one might expect these to become independent from the specific set $\bm{X}^{(j)}$ at the limit $n^{(j)}_p\rightarrow \infty$. The same is true for the inner product $\bm{\Phi}^T_{(j)}\bm{u}^{(j)}$ in \eqref{w_bar}.

Therefore, assuming that each snapshot is sufficiently dense, we approximate:
\begin{equation}
\label{eq:A_approximation}
\bm{\Phi}^T_{(j)}\bm{\Phi}_{(j)}\approx \frac{1}{n_t} \bm{\Phi}_E^T\bm{\Phi}_E\,, 
\end{equation}
and thus use \eqref{corr_f2} to write \eqref{light} as: 
\begin{equation}
\label{eq_Simp}
\langle\bm{W}\rangle\approx \bm{W}_E = \bigl (\bm{\Phi}_E^T\bm{\Phi}_E  +\alpha \bm{I}\bigr)^{-1}\bm{\Phi}^T_E \bm{U}_E \,.
\end{equation}

With the help of \eqref{eq_Simp}, we can therefore compute the mean of a random field through a single RBF regression of the ensemble of points. {The approach uses ``meshless'' collocation points (see \cite{Zhang2000, Chen2014, Fornberg2015}) because it does not require the definition of a computational mesh (with nodes, elements and connectivity) to compute derivatives. It is ``binless'' because the spatial distribution of flow statistics are regressed globally and not computed in local bins.}





\section{Numerical Recipes}\label{sec:3}

This section describes the numerical details in the implementation of the RBF regression described in the previous section. Subsection \ref{sec:3p1} reviews the methods to introduce physics-based constraints while subsection \ref{sec:3p2} describes the Partition of Unity Method (PUM) to minimize the memory requirements.

\vspace{-3mm}
\subsection{Constrained RBFs}\label{sec:3p1}

The RBF regression in \eqref{RBF_SOL} can be constrained using Lagrange multipliers and the Karush–Kuhn–Tucker (KKT) condition as shown in \cite{Sperotto2022a}.

The current implementation in SPICY \citep{Sperotto2024} allows to set linear constraints and quadratic penalties. These are used to impose or to penalize the violation of linear constraints such as Dirichlet and Neumann conditions, as well as divergence-free or curl-free conditions. Following the notation in \eqref{RBF_SOL}, the weight vector defining the RBF regression of the data ($\bm{X}_*,\bm{U}_*$) minimizes the following augmented cost function:
\begin{equation}
    \begin{split}
        J^\star (\mathbf{w}, \bm{\lambda}) &= \vert\vert \bm{U}_* - \bm{\Phi}(\bm{X}_*) \bm{W} \vert\vert_2^2 \\
        &+ \bm{\lambda}^T(\bm{\mathcal{L}}(\bm{X}_\mathcal{L}) \bm{W} - \bm{c_\mathcal{L}})\\
        &+ \alpha_\nabla \vert\vert \bm{D}_\nabla(\bm{X}_g) \bm{W} \vert\vert_2^2.
        \label{eq:cost_function}
    \end{split}
\end{equation}

{The first term is the least squares error. A minimization solely focused on this term yields the unconstrained solution in \eqref{RBF_SOL}. The second term is related to \emph{hard} linear constraints. The vector $\bm{\lambda}$ collects the associated Lagrange multipliers: these are additional unknowns to be identified in the constrained regression. The reader is referred to \cite{Sperotto2022a} for more details on the shape and formation of these matrices.}

The third term in \eqref{eq:cost_function} is a quadratic penalty, which in this work is solely used to penalize violations of the divergence-free condition, set on the full set of $n_g$ points with coordinates $\bm{X}_g$. The importance of the penalty is controlled by the parameter $\alpha_{\nabla}\in\mathbb{R}^+$. Penalties are \emph{soft constraints}: they promote but do not enforce a condition and require the a-priori (and not trivial) definition of $\alpha_\nabla$. On the other hand, their implementation is computationally much cheaper because penalties bring no new unknowns. The current implementation allows both constraints and penalties, to offer a compromise between the strength of hard constraints and the limited cost of penalties.

The problem of minimizing \eqref{eq:cost_function} can be cast into the problem of solving a linear system of the form:
\begin{equation}
    \label{eq:linear_system}
    \begin{pmatrix}
        \bm{A} & \bm{B}\\
        \bm{B}^T & \mathbf{0}
    \end{pmatrix}
    \begin{pmatrix}
        \bm{W}\\
        \bm{\lambda}
    \end{pmatrix} =
    \begin{pmatrix}
        \bm{b}_1\\
        \bm{b}_2
    \end{pmatrix},
\end{equation} {where $\bm{A} \in \mathbb{R}^{3 n_b \times 3 n_b}$ is computed from the training and penalty points, $\bm{B} \in \mathbb{R}^{3 n_b \times n_\lambda}$ is computed from the linear constraints, $\bm{b}_1$ is associated with the training data, and $\bm{b}_2$ is associated with the constraints. The vector $\bm{\lambda} \in \mathbb{R}^{n_\lambda}$ gathers the Lagrange multipliers for which the system must also be solved. The reader is referred to \cite{Sperotto2022a} for details on the matrices and efficient numerical methods for the system solution.} {It is worth stressing that this work solely considered equality constraints (e.g. divergence free of the mean flow field), although inequality constraints (e.g. positiveness of the Reynolds stresses) could also be included. These require the solution of a quadratic programming problem \citep{Boyd2004, Nocedal2006} and are currently under investigation.}

In what follows, we introduce the notation $\widetilde{\bm{U}}(\mathbf{x})=\bm{\Phi}(\mathbf{x})\bm{W}=\mbox{RBF}(\bm{U}_*,\bm{X}_*)$ to refer to the analytic approximation obtained by solving the constrained regression \eqref{eq:linear_system} for the training data $(\bm{X}^*,\bm{U}^*)$.

\vspace{-3mm}
\subsection{The Partition of Unity Method (PUM)}\label{sec:3p2}

An important limitation of the constrained RBF framework is the large memory demand due to the large size and the dense nature of the matrices involved in \eqref{eq:linear_system}. This problem can be mitigated using {compact support bases to make the system sparse and accessible to iterative methods for sparse systems or} the Partition of Unity Method (PUM) {to divide the problem into smaller blocks and enable direct solvers}. {We leave a detailed comparison (or possible combination) of the two approaches for future works, and here focus on the second because a preliminary investigation showed that it was faster and generally more accurate}.

The PUM was proposed by \cite{Melenk1996} in the context of the Finite Element Method, explored for interpolation purposes in \cite{Wendland2002, Cavoretto2021} and extensively developed by \cite{Larsson2013,Larsson2017,Cavoretto2019,Cavoretto2020} for the meshless integration of PDEs. The general idea of RBF-PUM is to split the regression problem in different portions (partitions) of the domain. Different PUM approaches have been proposed; these could be classified into ``global'' or ``local''. A global approach solves one large regression problem (e.g. \cite{Larsson2017}) which is made significantly sparser by the partitioning. A local approach solves many smaller regression problems (e.g. \cite{demarchi}) treating the regression in each portion as independent from the other.

In the context of data assimilation for image velocimery, the RBF-PUM has been recently used in \cite{Li2021} for smooth gradient computation and in \cite{Ratz2022b} for super-resolution. Recently, the extension of the RBF-PUM to include constraints has been proposed in \cite{Li2024}, following the stable gradient computation by \cite{Larsson2013}, and combined with a Lagrangian tracking approach. {Our approach differs from \cite{Li2024}'s in that we use a heuristic treatment of the derivatives at the intersection of the patches, which we found to be more stable.}

\begin{figure}
    \centering
    \includegraphics[width=0.49\textwidth]{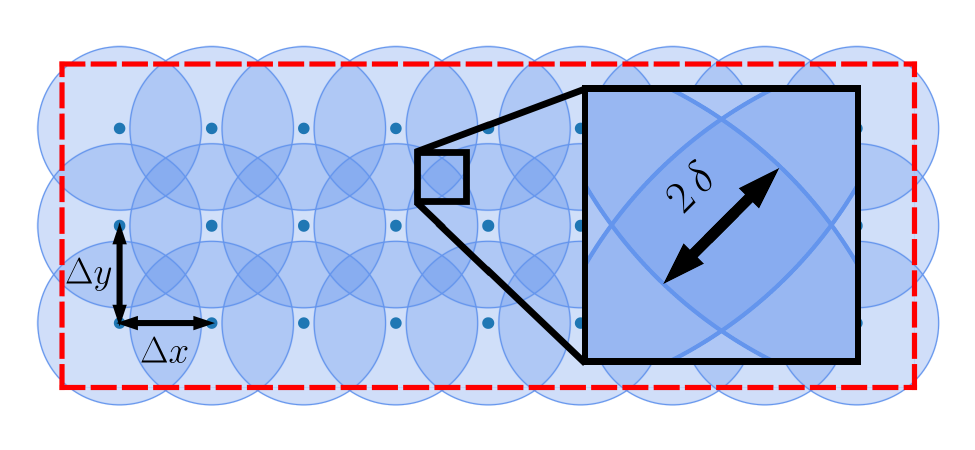}
    \caption{Example of a domain $\chi$ (dashed red lines) being covered by a total of 27 circular patches (blue circles) on a regularly spaced grid of $\Delta x$, $\Delta y$. The zoom-in (black solid lines) on the right hand side shows the overlap $\delta$ between patches}
    \label{fig:PUM_patches}
\end{figure}

To illustrate the proposed approach we first briefly recall the PUM with the help of Figure \ref{fig:PUM_patches}. The measurement domain $\chi$ is covered by $M$ spherical patches $\chi_m$ such that $\bigcup_{m=1}^{M} \chi_m \supset \chi$. In the 2D example of Figure \ref{fig:PUM_patches}, the rectangular domain (red dashed line) is covered by 27 patches (blue circles) with a regular spacing $\Delta x$ and $\Delta y$. The minimum radius to cover the entire domain is $r' = \sqrt{\Delta x^2 + \Delta y^2} / \sqrt{2}$. However, following \cite{Larsson2017}, the regression performs better if patches are partially overlapping, that is if the radius $r'$ is stretched by a factor $\delta$ to $r = r' (1 + \delta)$. This radius is used for every patch $\chi_m$. The overlap is visualized in the zoom-in (black solid lines) of Figure \ref{fig:PUM_patches}.

A weight function $\Omega_{(m)}$ is assigned to each patch. This function merges the contributions from the overlapping patches and is constructed such that:
\begin{equation}
    \sum_{m=1}^{M} \Omega_{(m)} ( \mathbf{x}) = 1, \quad \forall \mathbf{x} \in \chi.
\end{equation}

The weight functions are generated by applying the method by \citet{Shepard1968} for compactly supported functions, which gives:
\begin{equation}
    \label{eq:patch_weight}
    \Omega_{(m)}(\mathbf{x}) = \frac{\psi_m(\mathbf{x};c_m)}{\sum_{q = 1}^M \psi_q (\mathbf{x};c_q) },
\end{equation} where $\psi_m(\bm{x})$ is a compactly supported generating function, centered on $c_m$ in the $m$-th patch. An example for such a generating function is the Wendland $C^2$ function \citep{Wendland1995} which is defined as:
\begin{equation}
    \begin{split}
        &\psi_m(\mathbf{x}\vert \bm{c}_m, r) = \\
        & \hspace{1.1cm} \left(4 \frac{\vert\vert \mathbf{x} - \bm{c}_m \vert\vert^2}{r} + 1\right)\left(1 - \frac{\vert\vert \mathbf{x} - \bm{c}_m \vert\vert^2}{r}\right)^4_+\,,
    \end{split}
    \label{eq:wendland}
\end{equation} where $r$ is the radius of the function and the subscript $_+$ is the positive part of a function, i.e. $(a)_+ = a \; {\,\mbox{if}\, a > 0}$ and $(a)_+ = 0 \; {\,\mbox{if}\, a < 0}$. 

The $M$ patches are used to identify $M$ portions of datasets, each contained within the area $\Omega_{(m)}(\mathbf{x})\neq 0$ with $m=1,\dots M$.
The partitioning can be interpreted as a partitioning of the linear system \eqref{eq:block_struct} and the augmented cost function \eqref{eq:cost_function}. The partitioning consists in multiplying both the data and the constraints by the local weight function. That is, given the full dataset $(\bm{X}_*,\bm{U}_*)$, the data used for the local (constrained) regression in patch $m$ is $\bm{U}_{(m)}=\Omega_m(\bm{X}_{*,m}) \bm{U}_{*,m}$ and the bases used in each patch is $\bm{\Phi}_{(m)}=\Omega_m(\bm{X}_*)\bm{\Phi}(\bm{X}_*,\bm{X}_{c,m})$, with $\bm{X}_{c,m}$ considering only the subset of collocation points inside the $m$-th patch. {Similarly, all linear constraint operators $\mathcal{L}(\bm{X})$ and their values $\bm{c}_\mathcal{L}$ in \eqref{eq:cost_function} and \eqref{eq:linear_system} are weighted by the weight function $\Omega(\bm{X}_\mathcal{L})$. Then, each local regression can be carried out solving the local linear system \eqref{eq:linear_system} to obtain the local weights $\bm{W}_m$. Finally, given the set of local sets of weights, the analytical expression over the full domain is:}
{\begin{equation}
\label{PUM_ANSATZ}
\widetilde{\bm{U}}(\mathbf{x})=\sum^{M}_{m=1} \Omega_m(\mathbf{x})\bm{\Phi}(\mathbf{x},\bm{X}_{c,m}) \bm{W}_m\,.
\end{equation}}

{To compute derivatives, we use a heuristic treatment that supersedes the product rule and sets all derivatives of the weight functions to zero. Therefore, the partial derivative along $x$, for example, reads:}
{\begin{equation}
    \partial_x \widetilde{\bm{U}}(\mathbf{x})=\sum^{M}_{m=1} \Omega_m(\mathbf{x}) \partial_x \bm{\Phi}(\mathbf{x},\bm{X}_{c,m}) \bm{W}_m\,.
\end{equation}}

\begin{figure}[t]
    \center
    \includegraphics[width=0.48\textwidth]{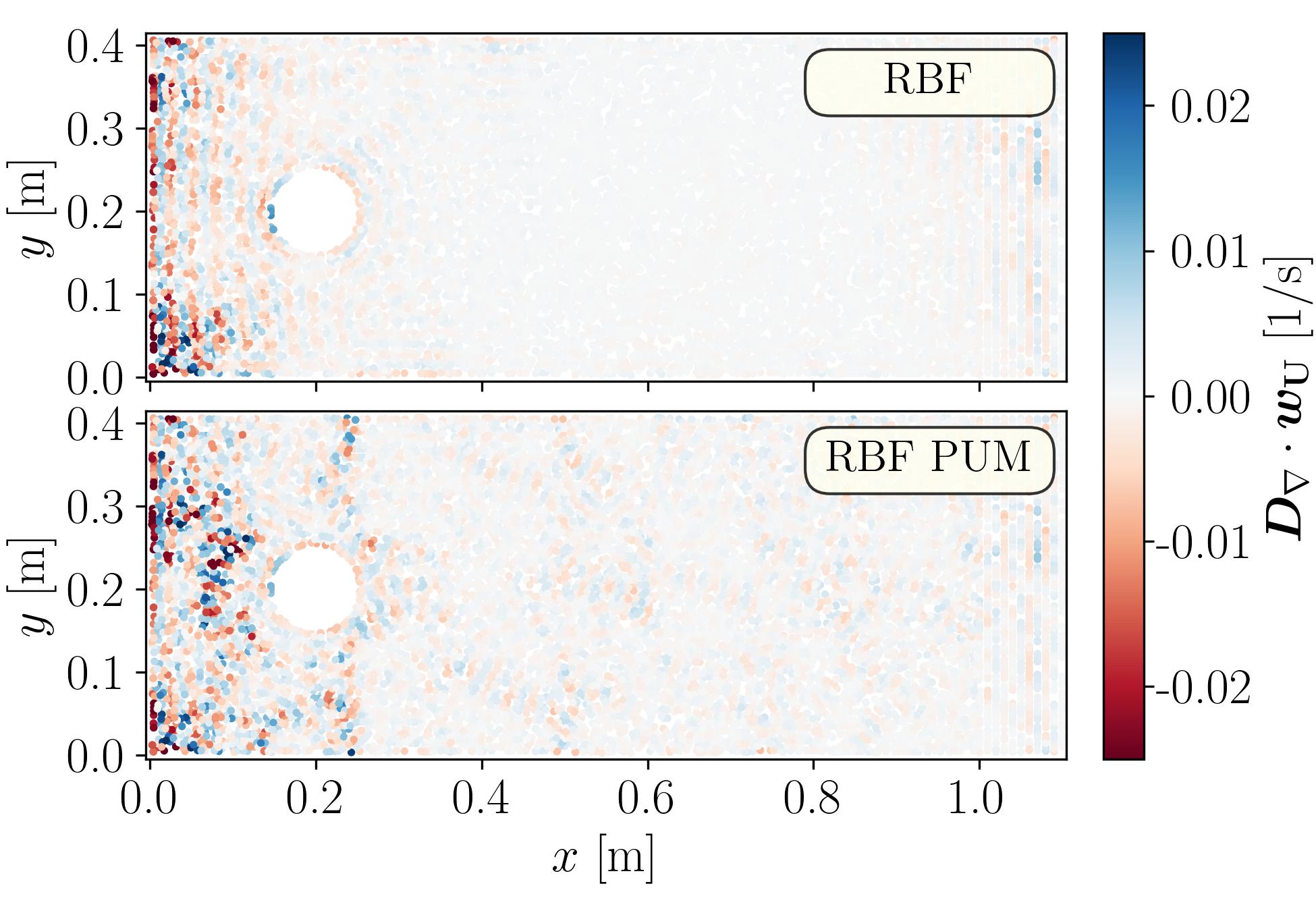}
    \caption{Divergence computed from the analytical RBF representation. Top: Constrained, global regression from \citet{Sperotto2022a}, Bottom: RBF-PUM with locally constrained regressions}
    \label{fig:pum_derivative_justification}
\end{figure}

To illustrate the performances of the PUM implementation, we consider the second test case in \citet{Sperotto2022a}, which is the regression of the flow past a cylinder in laminar conditions. We compare both our local PUM with a classic, global RBF regression. Figure \ref{fig:pum_derivative_justification} shows the analytical divergence field of the standard RBF regression at the top and the one of RBF-PUM at the bottom. {Both use solenoidal and Dirichlet constraints on the boundaries as well as a divergence-free penalty in every training point.} The largest differences are at the inlet and close to the cylinder where the gradients are largest. The magnitudes are comparable, and no pattern of the patches is visible. A comparison of the mean flow (not shown here) likewise only shows minor differences. The computational time of the RBF-PUM is an order of magnitude shorter. Further gains are possible by solving each of these $M$ problems in parallel on multiple processors, but we leave these developments to future improvement. In its current implementation, the PUM allowed to process millions of vectors on a modest laptop with 8GM RAM.

\begin{figure*}[htbp]
    \center
    \includegraphics[width=0.92\textwidth]{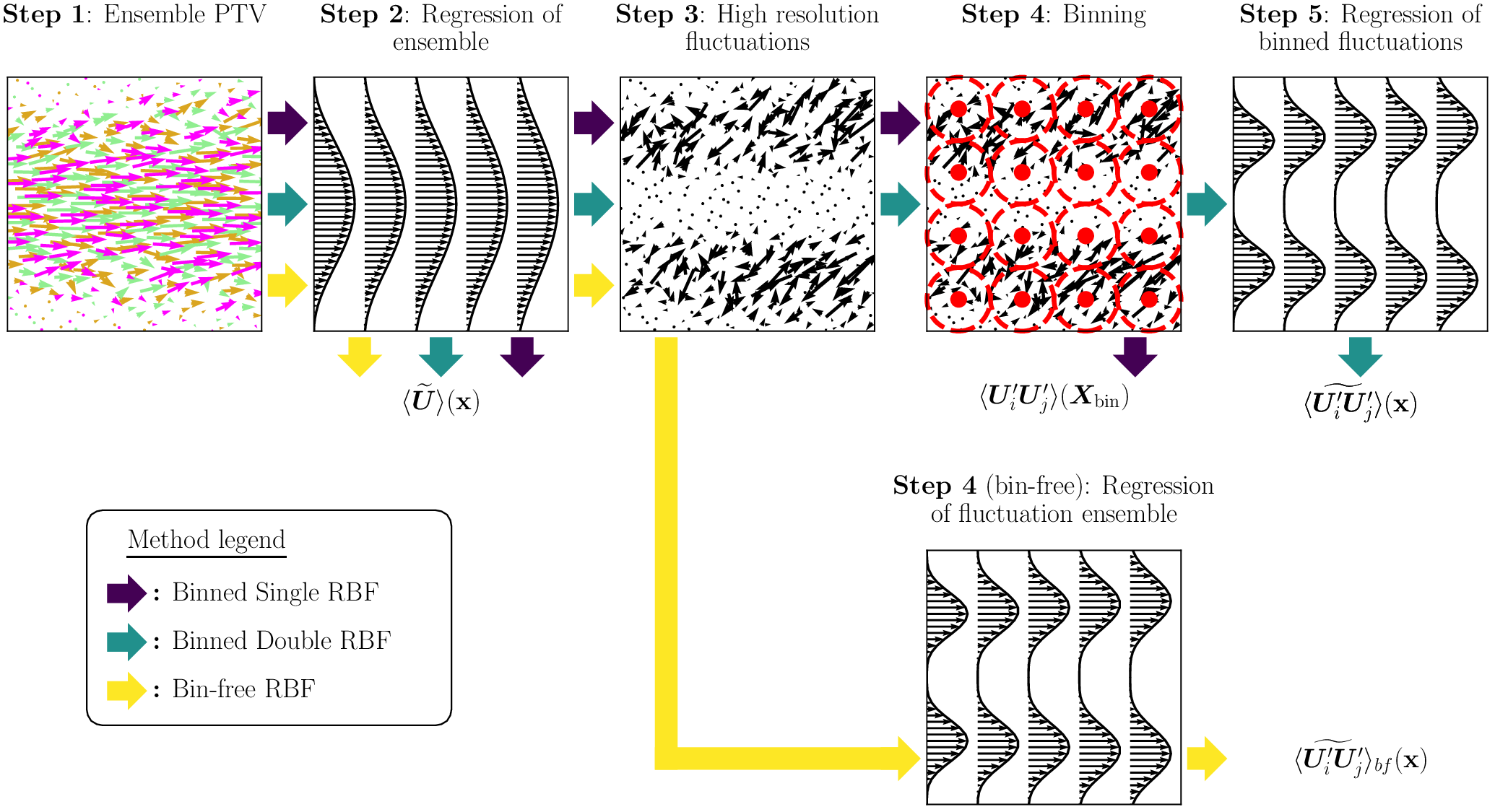}
    \caption{Flowchart explaining the processing pipeline of the three proposed RBF-based methods. The colors of the arrows correspond to each of the three methods according to the legend. All three methods subtract the global mean field analytically and then extract higher order statistics using (1) binning (Binned Single RBFs, purple) (2) binning and RBF regression (Binned Double RBF, teal) or (3) only an RBF regression (Bin-free Double RBF, yellow)}
    \label{fig:method_flowchart}
\end{figure*}

\section{Selected Algorithms for benchmarking}\label{sec:5}

\subsection{Traditional binning approaches}\label{sec:5p1}

We consider two traditional binning methods, namely the Gaussian weighting by \citet{Aguei1987} and the polynomial fitting by \citet{Agueera2016}. These are described in Section \ref{ss4p1p1} and \ref{ss4p1p2} respectively. These have in common that none of the statistical quantities are expressed as continuous functions. The statistics are only available at the bin's center, and higher resolution and gradients can only be obtained through further processing. We do not consider the top-hat approach since its shortcomings are well-known \citep{Agueera2016}. While all methods can extract higher-order statistics, we restrict our descriptions to first and second-order statistics for velocity fields, i.e. the mean flow and Reynolds stresses.

\subsubsection{Gaussian weighting}\label{ss4p1p1}

The Gaussian weighting \citep{Aguei1987} tackles unresolved velocity gradients by weighting points in every bin with a Gaussian. This simple approach gives less impact to points far from the bin center, mitigating the effects of unresolved mean flow gradients. However, weighting reduces the effective number of samples and thus decreases statistical convergence. In this work, we choose a standard deviation of $D_b/3$ for the Gaussian weighting functions, with $D_b$ the bin diameter.

\subsubsection{Polynomial fitting}\label{ss4p1p2}
The local polynomial fitting of \citet{Agueera2016} fits the ensemble fields within a bin with a polynomial function up to second order, providing a continuous function of the \textit{local} mean flow. This continuous function is used for two purposes. First, it is evaluated in the bin center to provide the mean velocity in the bin. Second, it is evaluated in all data points within a bin, and subtracted to the instantaneous velocities to compute the velocity fluctuations. Higher order statistics are sampled on the mean-subtracted fields through a top-hat-like approach.

\vspace{-3mm}

\subsection{RBF-based approaches}\label{sec:5p2}

The RBF approaches build on the mathematical background introduced in Sections \ref{sec:2} and \ref{sec:3}, and in particular on the assumption that the expectation operator can be approximated by a regression in space. The framework was implemented with three variants in three algorithms, named `Binned Single RBF', `Binned Double RBF', and `Bin-free RBF'. These algorithms share several common steps, which are recalled in the flowchart in Figure \ref{fig:method_flowchart}. The sequence of steps for each method is traced using arrows of different colors, recalled in the legend on the bottom left.

\begin{itemize}
\item \textbf{Step 1}. The starting point for all methods is an ensemble flow field that is assumed to have gathered enough realizations to provide statistical convergence.  This is indicated in Figure \ref{fig:method_flowchart}, using different colors for fields in different snapshots.

\item \textbf{Step 2}. For all methods, the mean flow is computed in the same way using a PUM-based constrained regression RBF of the ensemble. This provides the analytical mean flow field:  
\begin{align}
\label{RBF_1}
    \langle \widetilde{\bm{U}} \rangle (\mathbf{x}) = \text{RBF}(\bm{X}_E, \bm{U}_E)\,.
\end{align}

\item \textbf{Step 3}. The function \eqref{RBF_1} is used to compute the ensemble of velocity fluctuations by subtracting the mean field $\langle \widetilde{\bm{U}} \rangle (\bm{X}_E)$ to the ensemble field:  
\begin{align}
    \bm{U}^\prime(\bm{X}_E) = \bm{U}_E - \langle \widetilde{\bm{U}} \rangle (\bm{X}_E)\,.
\end{align}

This field is then used to compute all the products $\bm{U}_i^\prime \bm{U}_j^\prime(\bm{X}_E)$, that are required by all methods in the following steps. This is the last common step for the three methods.

\vspace{3mm}

\textbf{Binned Single RBF}

\item \textbf{Step 4}. This method now interrogates the ensemble fields of products $\bm{U}_i^\prime \bm{U}_j^\prime(\bm{X}_E)$ with a standard binning process. This is the simplest approach and most similar to the one of \citet{Agueera2016}, with the only difference being a globally smooth physics constrained regression instead of a local (locally smooth) polynomial regression. 
The binning process yields a discrete field of second order statistics on the binning grid $\bm{X}_\text{bin}$, i.e. $\langle\bm{U}_i^\prime \bm{U}_j^\prime \rangle (\bm{X}_\text{bin})$.

\vspace{3mm}

\textbf{Binned Double RBF}
\vspace{1mm}

\item \textbf{Step 5}. This method builds on the binning grid from \textbf{Step 4} of Binned Single RBF with a second regression:
\begin{equation}
\label{bin_double}
    \langle \widetilde{\bm{U}_i^\prime \bm{U}_j^\prime} \rangle (\mathbf{x}) = \text{RBF}( \bm{X}_\text{bin}, \langle \bm{U}_i^\prime \bm{U}_j^\prime \rangle (\bm{X}_\text{bin}))\,.
\end{equation}

This regression has two purposes. First, it gives an analytical expression for not only the mean but also the Reynolds stresses. Second, it smoothes noisy Reynolds stress fields which occur if the number of samples within a bin is insufficient for convergence. Therefore, fewer samples are required in experiments.

\vspace{3mm}

\textbf{Bin-Free RBF}
\vspace{1mm}

\item \textbf{Step 4} (bin-free). The bin-free approach deviates from the former two methods after \textbf{Step 3}. This method works on the ensemble fields of products $\bm{U}_i^\prime \bm{U}_j^\prime(\bm{X}_E)$ without binning, replacing the ensemble operators with the RBF (spatial) regression of the ensemble: 
\begin{align}
\label{bin_free}
    \langle \widetilde{\bm{U}_i^\prime \bm{U}_j^\prime} \rangle_{bf} (\mathbf{x}) = \text{RBF}(\bm{X}_E, \bm{U}_i^\prime \bm{U}_j^\prime(\bm{X}_E))\,,
\end{align} where the subscript $'bf'$ is used to distinguish the output of \eqref{bin_free} from the output in \eqref{bin_double}. The main advantage with respect to the previous approach is to by-pass the averaging effects of the binning. However, the computational cost and the complexity of the algorithm is higher, because the number of ensemble points in \eqref{bin_free} is larger than the number of bins in \eqref{bin_double}. Yet, if the same collocation points and shape parameters are reused, computations can be shared for the two successive regressions of bin-free RBF. 
\end{itemize}


\section{Selected test cases}\label{sec:4}
\vspace{-3mm}
\subsection{1D Gaussian process}\label{sec:4p1}

A synthetic 1D test case was designed to illustrate the relevance of the assumption that the average of multiple regressions can be approximated by a single regression of the ensemble (see Section \ref{sec:2p2}).

The 1D dataset is generated by sampling a 1D Gaussian process with average:
\begin{equation}
    u(x) = x + \frac{1}{6} \sin\left(\frac{3 \pi}{2} x\right),
\end{equation}
and covariance function:
\begin{equation}
\kappa(x_1,x_2)=\sigma_f \exp(-\gamma (x_2-x_1)^2)\,,
\end{equation} with $\gamma=12.5$ and $\sigma_f=0.01$. In a Gaussian process, the covariance function acts as a kernel function measuring the ``similarity'' between two points.

\begin{figure}[ht]
    \centering
    \includegraphics[width=0.47\textwidth]{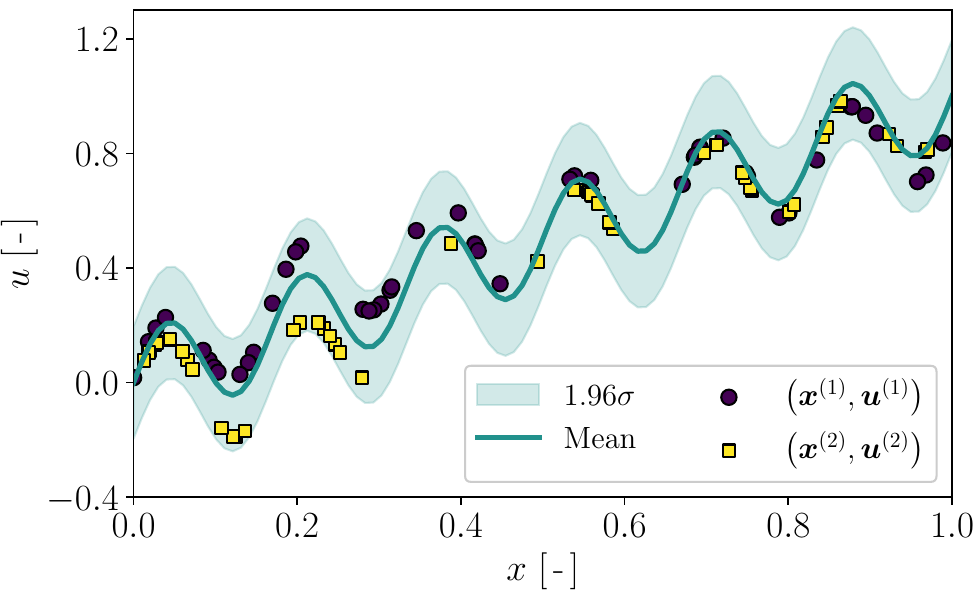}
    \caption{Test case 1: a 1D Gaussian process. The mean value is shown with a solid line and the \SI{95}{\percent} interval is shown by the shaded area. The scattered markers represent two different members of the ensemble $\left(\bm{x}^{(1)}, \bm{u}^{(1)}\right)$ and $\left(\bm{x}^{(2)}, \bm{u}^{(2)}\right)$) with $n_p^{(j)} = 50$}
    \label{fig:gaussian_process_target_function}
\end{figure}

The domain $x$ extends from 0 to 1 and total of $n_E$ ensembles with $n_s$ samples are sampled from this process. Figure \ref{fig:gaussian_process_target_function} shows two members of the ensembles $\bigl(\bm{x}^{(1)}, \bm{u}^{(1)}\bigr)$ and $\bigl(\bm{x}^{(2)}, \bm{u}^{(2)}\bigr)$ together with the process average and the $95\%$ confidence interval in shaded area. We verify the validity of assumption \eqref{eq_Simp} by varying the size of the ensemble and the sample size.

\subsection{3D Synthetic turbulent jet}\label{sec:4p2}
The second synthetic test case is a three-dimensional, jet-like, turbulent velocity field. This is used to compare the proposed RBF-based methods with classic binning approaches on a case for which the ground truth is available. The synthetic test case is set up in the domain $(x,y,z)\in [-100, 100] \times [-75, 75] \times [-75, 75]\;$voxels ($\mbox{vox}$). Using cylindrical coordinates $\mathbf{u} = (u_x,\, u_r,\, u_\theta)$, the mean flow has axial component given by:
\begin{equation}
    \langle u_x \rangle (x, r) = \frac{U_0}{2} \left[ 1 + \text{cos}\left(\frac{2 \pi r}{\lambda(x)}\right) \right],
\end{equation}
where $U_0 = 3\;$vox is the maximum displacement, $r = \sqrt{y^2 + z^2}$ is the radius and $\lambda(x)$ defines the width of the profile which increases linearly from 60 to 90\;vox. The mean velocity field is zero in the other components, i.e. $\langle u_r \rangle=\langle u_\theta \rangle =0$. Therefore, this field is not divergence-free and is solely used for demonstration purposes. 

Synthetic turbulence is added in a ring with Gaussian noise. The synthetic shear layer is located at $r = 0.4 \lambda(x)$ with a width of $0.5 \lambda(x)$ corresponding to a standard deviation:
\begin{equation}
    \sigma_N (x, r) = \frac{3}{2 \sqrt{10}}\left[ 1 + \text{cos}\left(\frac{2 \pi (r - 0.4 \lambda(x))}{0.5\lambda(x)}\right) \right].
\end{equation}

This is used to construct the velocity fluctuations $u^{\prime}_x$, $u^{\prime}_r$ and $u^{\prime}_\theta$ as a multivariate Gaussian $\bm{u}^\prime (\mathbf{x})\sim \mathcal{N}(\bm{\mu},\bm{\Sigma})\in\mathbb{R}^{3}$ with mean $\bm{\mu}$ and covariance matrix $\bm{\Sigma}$ defined as:

\begin{equation}
\mathbf{u}^\prime(\mathbf{x})=
\begin{pmatrix}
    u_x^\prime \\
    u_r^\prime \\
    u_\theta^\prime
\end{pmatrix} \sim 
 \mathcal{N}\left( \begin{pmatrix}
    0 \\
    0 \\
    0
\end{pmatrix}, \begin{bmatrix}
    \sigma_N^2 & 0.7 \,\sigma_N^2 & 0 \\
    0.7\,\sigma_N^2 & \sigma_N^2 & 0 \\
    0 & 0 & \sigma_N^2
\end{bmatrix} \right)\,.
\end{equation}

\begin{figure}[t]
    \center
    \includegraphics[width=0.5\textwidth]{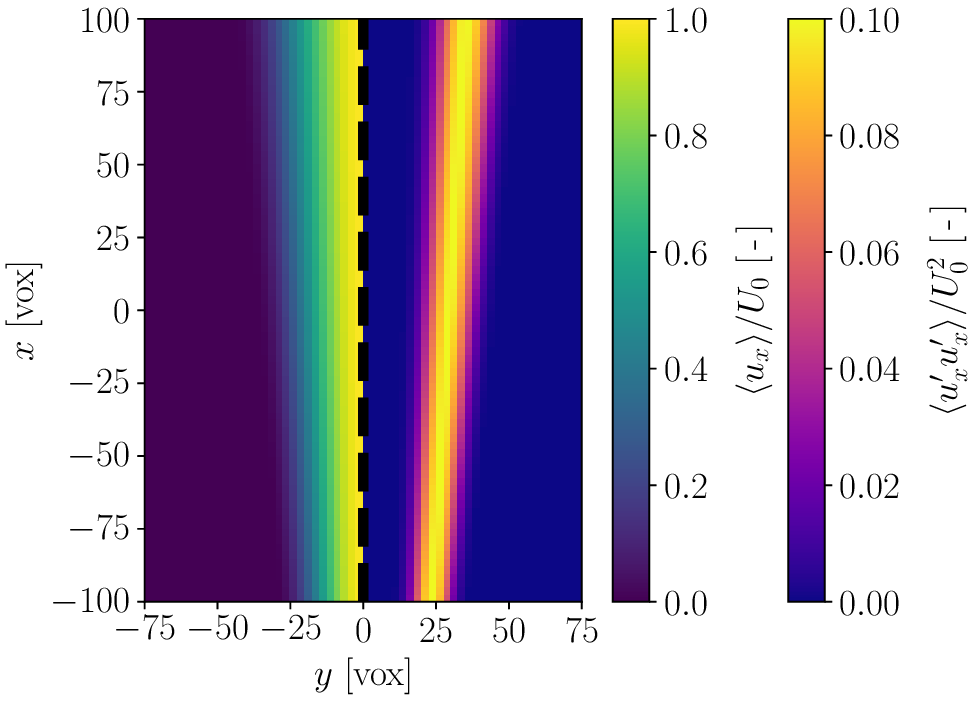}
    \caption{Test case 2. Exact velocity field of the synthetic jet at $z = 0$\,vox. Axial mean flow (left) and axial normal Reynolds stress (right)}
    \label{fig:case2_groundtruth}
\end{figure}

That is, the fluctuations $u^{\prime}_x$ and $u^{\prime}_r$ are correlated while the fluctuation $u^{\prime}_\theta$ is not.
Figure \ref{fig:case2_groundtruth} shows the contour map of the axial mean flow (on the left) and the axial fluctuation $u^{\prime}_x$ (on the right).

A total of $1 \cdot 10^6$. scattered random points were taken as the velocity field ensemble. We further contaminate these ideal fluctuations by adding uniform noise according to 
$\mathbf{u}_n(\mathbf{x}) = \mathbf{u}(\mathbf{x}) (1 + \mathbf{q}(\mathbf{x}))$. Here $\mathbf{q}(\mathbf{x}) = (q_x(\mathbf{x}), q_y(\mathbf{x}), q_z(\mathbf{x}))$ is a noise vector for each velocity component, where each component is independently sampled from a rectangular distribution in the interval $[-0.1, 0.1]$.

\subsection{3D Experimental turbulent jet}\label{sec:4p3}
\begin{figure*}[htbp]
    \center
    \includegraphics[width=0.98\textwidth]{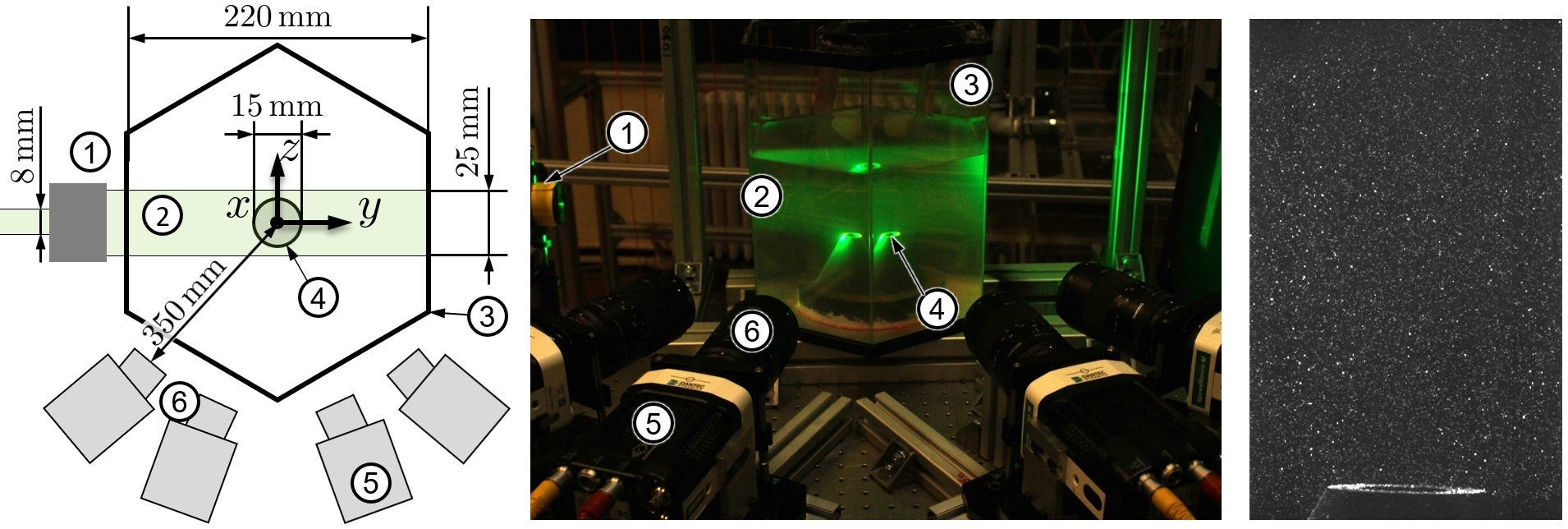}\\
    \hspace{1.3cm}\textbf{(a)} \hspace{6.00cm} \textbf{(b)} \hspace{5.05cm} \textbf{(c)}\hspace{-0.05cm}
    \caption{\textbf{(a)} Top-down sketch of the experimental facility with the right-handed coordinate system and \textbf{(b)} image of the facility during the acquisition. Laser light (1) enters the top-hat illumination optics to produce a  volumetric illumination (2) which enters the hexagonal tank (3). The illumination is centered above the jet nozzle (4) which is located at the bottom of the tank. Four high-speed cameras (5) with \SI{100}{\milli\meter} objectives (6) record the jet in an arc that covers approximately \ang{125}; \textbf{(c)} example of an acquired raw image for the left-most camera}
    \label{fig:experimental_setup}
\end{figure*}

The third test case is a 3D PTV measurement of an underwater jet at the von Karman Institute. The setup of the facility is sketched on the left-hand side of Figure \ref{fig:experimental_setup}, with a picture of the facility in the center of the Figure. The jet nozzle with a diameter of $D = \SI{15}{\milli\meter}$ was located at the bottom of a hexagonal water tank with a width of \SI{220}{\milli\meter} and a free surface. The nozzle was fixed at the bottom of the tank and the origin of the coordinate system was set to the center of the nozzle exit. A centrifugal pump was connected to the back of the nozzle with a tube. The effects of the resulting Dean vortices were suppressed by installing a grid with a size of \SI{2}{\milli\meter} inside the nozzle. The inlet length from the grid to the exit of the nozzle was approximately $4D$ due to spatial constraints. The exit velocity $U_0$ of the jet was approximately \SI{0.45}{\meter\per\second}, which resulted in a diameter-based Reynolds number of \num{6750}.

\begin{table}[t]
    \centering
    \caption{Parameters of the experimental setup}
    \vspace{0.1cm}
    \begin{tabularx}{\columnwidth}{@{}p{0.52\columnwidth}p{0.45\columnwidth}@{}}
    \hline \\[-7pt]
    Nozzle diameter $D$ & \SI{15}{\milli\meter} \\
    Central jet velocity $U_0$ & \SI{0.45}{\meter\per\second} \\
    Reynolds number $\text{Re}_D$ & \num{6750} \\
    Medium & Water \\
    Camera type & Speedsense M310\\
    Number of cameras $N_\text{cam}$ & 4 \\
    Acquisition frequency $f_\text{acq}$ & \SI{1000}{\hertz} \\
    Camera resolution& \numproduct{1280 x 800}\;\si{\pixel} \\
    Camera exposure time $t_\text{exp}$ & \SI{250}{\micro\meter} \\
    Scaling factor & 14.7\;vox/mm \\
    Illum. vol. $(x\times y\times z)$ & \qtyproduct{67.5 x 40 x 22.5}{\milli\meter} \\
    & $(4.5D \times 2.5D \times 1.5 D)$ \\
    Lenses &  Samyang Macro \\
    & F2.8/\SI{100}{\milli\meter}\\
    Lens aperture $f_\#$ & 11 \\
    Scheimpflug adapter & Single axis \\
    Illumination & Quantronix Darwin-Duo \\
    & 527-80-M laser\\
    Seeding & Fluorescent microspheres \\
    Seeding diameter $d_p$ & \qtyrange{45}{53}{\micro\meter} \\
    \# tracked particles $n_p$ & \numrange{4000}{7000}\\
    Seeding density on sensor $N_\text{ppp}$ & 0.018\,ppp \\ [3pt]
    \hline
    \end{tabularx}
    \label{tab:experimental_parameters}
\end{table}

The flow was illuminated with a Quantronix Darwin Duo 527-80-M laser with a wavelength of \SI{527}{\nano\meter} and \SI{25}{\milli\joule} per pulse. The volumetric illumination was achieved with top-hat illumination optics from Dantec Dynamics and entered through the side of the tank. The optics produced a beam with a rectangular cross-section with an aspect ratio of $5:1$, which resulted in an illuminated volume of $(x \times y \times z) \approx 4.5D \times 2.5D \times 1.5 D$. The resulting scaling factor was approximately \SI{14.7}{\voxel\per\milli\meter}. Red fluorescent microspheres with a diameter ranging from \qtyrange{45}{53}{\micro\meter} and a density of \SI{1200}{\kilo\gram\per\meter\cubed} were used as tracer particles. The higher density of the particles allows to vary the seeding concentration by leveraging sedimentation over time. This is particularly helpful for the calibration refinement, which requires much lower seeding concentration ($0.005$\;ppp) than what used during the experiments ($0.018$\;ppp).

The density mismatch was not considered critical to the experiments, since the particles have a terminal velocity of approximately $u_T=\SI{0.25}{\milli\meter\per\second}$, that is about a thousandth of the free jet velocity in the free stream. Moreover, the Stokes number was small enough at $\text{Stk} \approx 5 \cdot 10^{-3}$ to have tracking errors below \SI{1}{\percent} \citep{Raffel2018}.

Four SpeedSense M310 high-speed cameras with a resolution of $1280 \times 800$\;\si{\pixel} were used to observe the flow in the region directly above the jet. The cameras had a distance of approximately \SI{350}{\milli\meter} from the jet center and were arranged in an arc of approximately \SI{125}{\degree} as is shown in to Figure \ref{fig:experimental_setup}. The cameras were equipped with Samyang Macro objectives (F2.8, $f = \SI{100}{\milli\meter}$, $f_\# = 11$) and long-pass filters to suppress the reflected laser light. All cameras were used in single-axis Scheimpflug arrangement with an angle of approximately 3 and \SI{12}{\degree} for the interior and exterior cameras, respectively. A total of 2000 time-resolved images were acquired with Dynamic Studio 8.0, at a frequency of \SI{1000}{\hertz}. This corresponds to a maximum displacement of \SI{8}{\voxel} for particles in the jet center. An example of an acquired raw image is displayed on the right-hand side of Figure \ref{fig:experimental_setup}, and Table \ref{tab:experimental_parameters} summarizes the experimental parameters.

The cameras were calibrated with a dotted calibration target (size \qtyproduct{100 x 100}{\milli\meter}, black dots on white background, diameter \SI{1.5}{\milli\meter}, pitch \SI{2.5}{\milli\meter}). The target was traversed in the range from $z = \pm \SI{15}{\milli\meter}$ throughout the volume by means of a translation stage with micrometric precision. Five images were acquired at equally spaced positions, and a 2nd-degree polynomial in all three axes was used as a calibration model. The resulting calibration error was approximately 0.15 and \SI{0.3}{\pixel} for the interior and exterior cameras, respectively. The calibration error was reduced using the procedure outlined by \citet{Bruecker2020}. For this, a total of 21 statistically independent images were recorded at a seeding concentration of approximately 0.005\;ppp.
After calibration refinement, the error of every camera was below \SI{0.05}{\pixel}.


The acquired images were processed with a mean subtraction over all images for each camera. Residual background noise was eliminated by clipping all pixels with an intensity below 60 counts. For each time step, the 3D voxel volume was reconstructed in a domain of approximately $L_x \times L_y \times L_z = 990 \times 550 \times 285$ in $x, y$, and $z$ using up to 10 iterations of the SMART algorithm \citep{Atkinson2009, Scarano2013}. 

\begin{figure*}[ht!]
    \centering
    \includegraphics[width=0.99\textwidth]{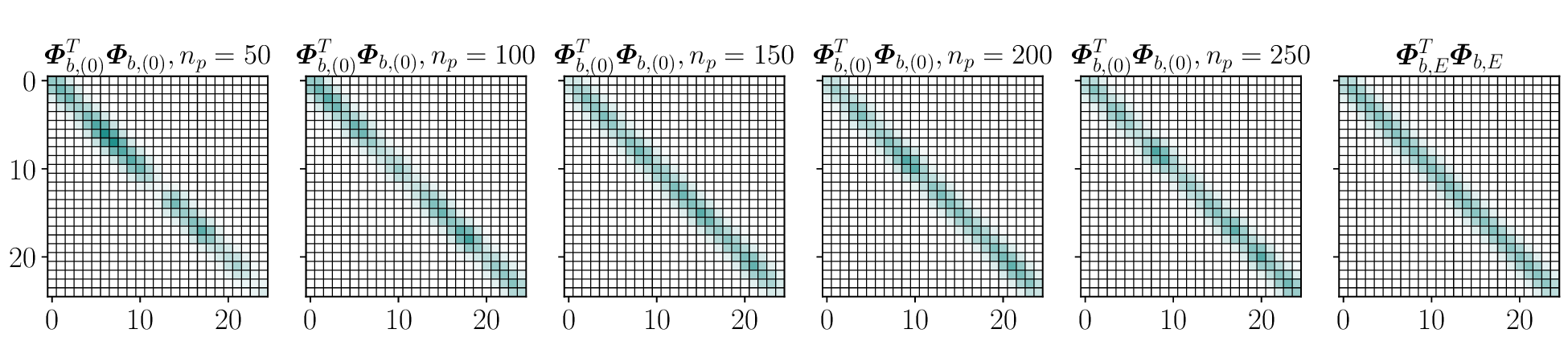}\\
    \hspace{0.45cm}\textbf{(a)} \hspace{2.2cm} \textbf{(b)} \hspace{2.2cm} \textbf{(c)}\hspace{2.3cm} \textbf{(d)}\hspace{2.3cm} \textbf{(e)}\hspace{2.3cm} \textbf{(f)}\hspace{-0.05cm}
    \caption{Test case 1: Matrix structure of $\bm{\Phi}_b^T \bm{\Phi}_b$. Subfigures \textbf{(a)}-\textbf{(e)} show the matrix for the five different sample sizes and subfigure \textbf{(f)} shows the matrix for the ensemble of points}
    \label{fig:case1_phi_matrices}
\end{figure*}

\begin{figure}[ht!]
    \centering
    \includegraphics[width=0.47\textwidth]{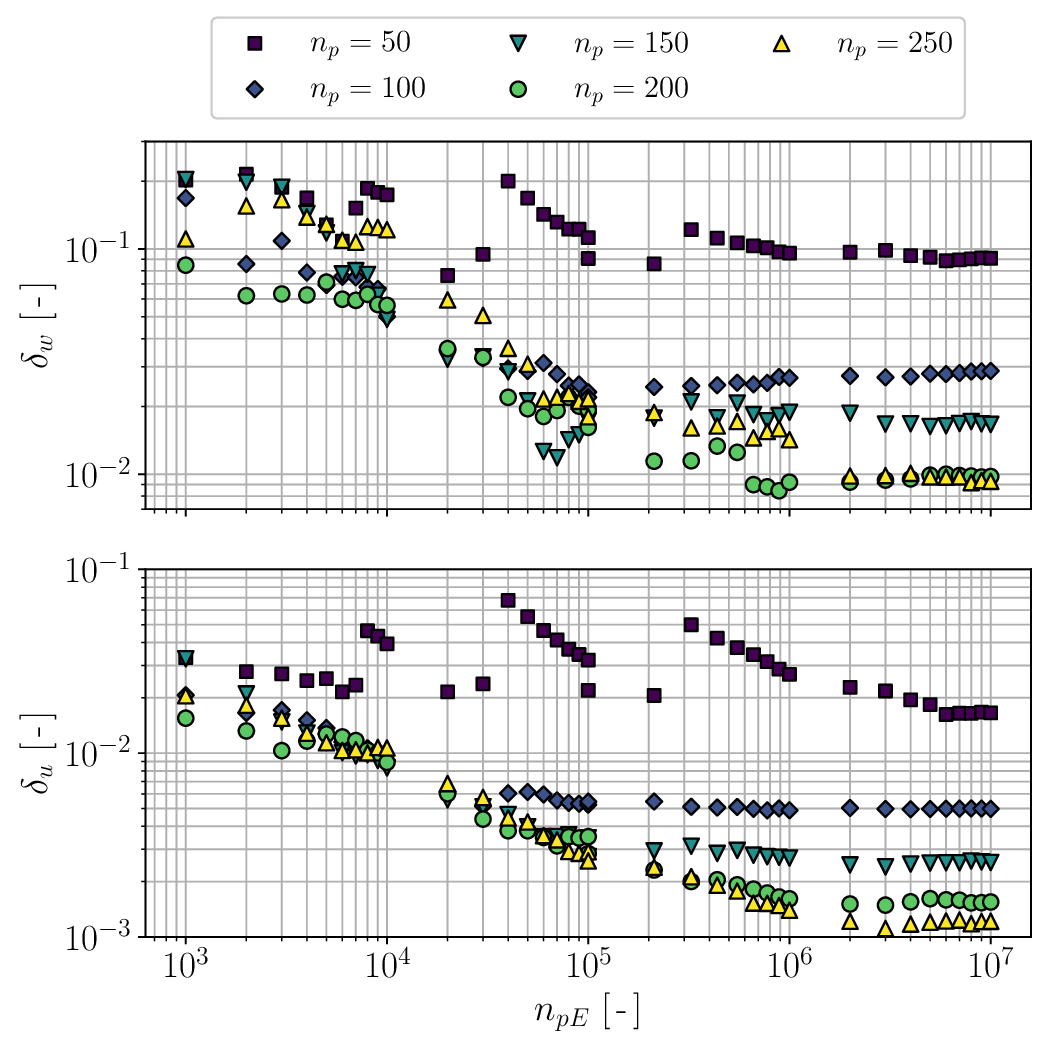}
    \caption{Test case 1: Convergence of the relative difference between the RBF regression of the ensemble and the ensemble of the RBF regressions. Difference in the weights (top) and the mean (bottom)}
    \label{fig:case1_convergence}
\end{figure}

For the given parameters, the fraction of ghost particles can be estimated according to \citet{Discetti2014}:
\begin{equation}
    \frac{N_\text{ghost}}{N_\text{true}} = N_\text{ppp} d_\tau L_z \left(1-e^{-N_s}\right)^{N_\text{cam}-2},
\end{equation} where $d_\tau = 2.5$\;px was the particle image diameter and $N_s = N_\text{ppp} \pi d_\tau^2/4$ the source density. It is important to highlight that the volume was not reconstructed in the full illumination depth of $1.5D$, but was reduced to $1.3D$ because of reduced intensity in the outer regions. The resulting \SI{10}{\percent} of ghost particles are treated through time-resolved information with predictors based on previous time steps. This increases the accuracy \citep{Malik1993, Cierpka2013} and allows to filter ghost particles which typically have a short track length \citep{Kitzhofer2009}.

After filtering out particles with a track length below 5 time steps, a total of \numrange{4000}{7000} vectors were computed at each snapshot. Three additional processing steps were applied. First, a normalized median test was used to remove outliers \citep{Westerweel2005}. Second, the domain depth was reduced to $1.1D$ because of an insufficient number of particles in the outer region, which negatively affected the RBF regression. Third, we only used data from every third time step, since this provides sufficient statistical convergence and a sufficient level of statistical independence of the snapshots in the shear layer. The resulting dataset consists of $3.35 \cdot 10^6$ particles in the ensemble used for the training.


\section{Results}\label{sec:6}

\subsection{1D Gaussian process}
\label{sec:6p1}

The main purpose of this illustrative test case was to compare the average of RBF regressions in \eqref{w_bar} with the RBF regression of the ensemble in \eqref{eq_Simp}. {In both cases, we use 25 evenly spaced RBFs with a radius of 0.06, which is defined as the distance at which the RBF reaches half its value. These values are chosen to sufficiently cover the domain and have a well-posed regression for the lowest seeding case. However, the lack of points leads to ill-conditioned matrices and thus, a strong regularization is needed}. The regularization parameter $\alpha$ in \eqref{RBF_SOL} was computed by setting an upper limit to the condition number $\kappa(\bm{H})$ of the matrix $\bm{H}=(\bm{\Phi}^T_*\bm{\Phi}_*)$ estimated as follows 

\begin{equation}
\kappa(\bm{H})\approx\frac{\lambda_M}{\alpha}\rightarrow \alpha=\frac{\lambda_M(\bm{H})}{\kappa_L}\,,
\end{equation} with $\lambda_M(\bm{H})$ the largest eigenvalue of $\bm{H}$ and $\kappa_L=10^{4}$ the upper limit of the condition number. {This regularization approach is used in all regressions in the remainder of this article, each with different values of $\kappa_L$.}

We consider a set of $n_{pE}$ samples in the ensemble, varying from $n_{pE} = 10^3$ to $10^7$. To compute the average of RBFs in \eqref{w_bar}, we assume that the ``snapshots'' from which each regression is carried out consists of $n_p$ samples, taken as $n_p = \{50, 100, 150, 200, 250\}$. Therefore, the number of regressions is $n_t=n_{pE}/n_p$: one could either work with many sparse snapshots (small $n_p$ and large $n_t$) or fewer dense snapshots (large $n_p$ and small $n_t$), but for the comparison with the ensemble approach, the same $n_{pE}$ is kept for all cases. The points are randomly sampled using a uniform distribution.

For each snapshot $i$, the regression evaluates the basis matrix $\bm{\varPhi}_{b,(i)}$ and computes the set of weights $\bm{w}^{(i)}$ using the unconstrained RBF regression in \eqref{RBF_SOL}, i.e. $\bm{w}^{(i)}=\mbox{RBF}(\bm{x}^{(i)},\bm{u}^{(i)})$.

Figure \ref{fig:case1_phi_matrices} compare the matrices $\bm{\varPhi}_{b,(0)}^T\bm{\varPhi}_{b,(0)}$ for snapshots with $n_p = \{50, 100, 150, 200, 250\}$ particles each together with the case using the full ensemble of points with $n_{pE}=10^7$. 
As expected, all matrices have a diagonal band proportional to the width of the RBFs. This is particularly smooth for the ensemble and shows ``holes'' for the sample matrices, which becomes more pronounced as $n_p$ is reduced. This is due to the uneven and overly sparse distribution of points in each sample. However, for sufficiently dense snapshots, it is clear that the all inner product matrices $\bm{\varPhi}_b^T\bm{\varPhi}_b$ converge to a prescribed function. This is the essence of the shift in paradigm from the ensemble averaging of regressions to the regression of the ensemble dataset $\bm{w}_E=\mbox{RBF}(\bm{x}_E,\bm{u}_E)$, with $(\bm{x}_E,\bm{u}_E)$ the ensemble dataset.

To analyze the impact of the sampling on the comparison between \eqref{w_bar} and \eqref{eq_Simp}, we define the $l_2$ discrepancy between the weights and the predictions as 
\begin{subequations}
\label{deltas}
    \begin{equation}
        \delta_{W} = \frac{\vert\vert \bm{w}_A - \bm{w}_E \vert\vert_2}{\vert\vert \bm{w}_A \vert\vert_2},\,
    \label{eq:error_norm}
    \end{equation}
    \begin{equation}
        \delta_{\langle u \rangle}= \frac{\vert\vert \langle \widetilde{\bm{u}} \rangle_A - \langle \widetilde{\bm{u}} \rangle_E \vert\vert_2}{\vert\vert \langle \widetilde{\bm{u}} \rangle_A \vert\vert_2},
    \end{equation}
\end{subequations} where $\bm{w}_A = \sum_{i=1}^{n_t} \bm{w}^{(i)} / n_t$, $\langle \widetilde{\bm{u}} \rangle_A = \sum_{i=0}^{n_t} \langle\widetilde{\bm{u}}^{(i)} \rangle / n_t$, with $\langle \widetilde{\bm{u}}^{(i)} \rangle=\bm{\Phi}_b(\bm{x}^{(i)})\bm{w}^{(i)}$ and $\langle \bm{u} \rangle_E=\bm{\Phi}_b(\bm{x}_{E})\bm{w}_{E}$.

Figure \ref{fig:case1_convergence} plots $\delta_w$ and $\delta_u$ in \eqref{deltas} as a function of the number of samples in the ensemble ($n_{pE}$) for the five choices of samples per snapshot $n_p$. The results shows that $n_p=50$ is clearly insufficient for the problem at hand. This is due to the fact that \eqref{eq:A_approximation} does not hold for most of the samples and an average of poor regressions is a poor regression.
However, as $n_p$ increases, convergence is observed with both $\delta$ dropping smoothly below \SI{1}{\percent} for $n_{pE}>10^4$ regardless of $n_p$. Moreover, this comparison shows that the discrepancies on the weight vectors are attenuated in the approximated solution. Although these results depend on the settings of the RBF regression, and in particular on the level of regularization, these results give a practical demonstration on the feasibility of approximating \eqref{w_bar} with \eqref{eq_Simp}.

\subsection{3D Synthetic turbulent jet}\label{sec:6p2}

\begin{figure*}[t]
    \center
    \includegraphics[width=0.98\textwidth]{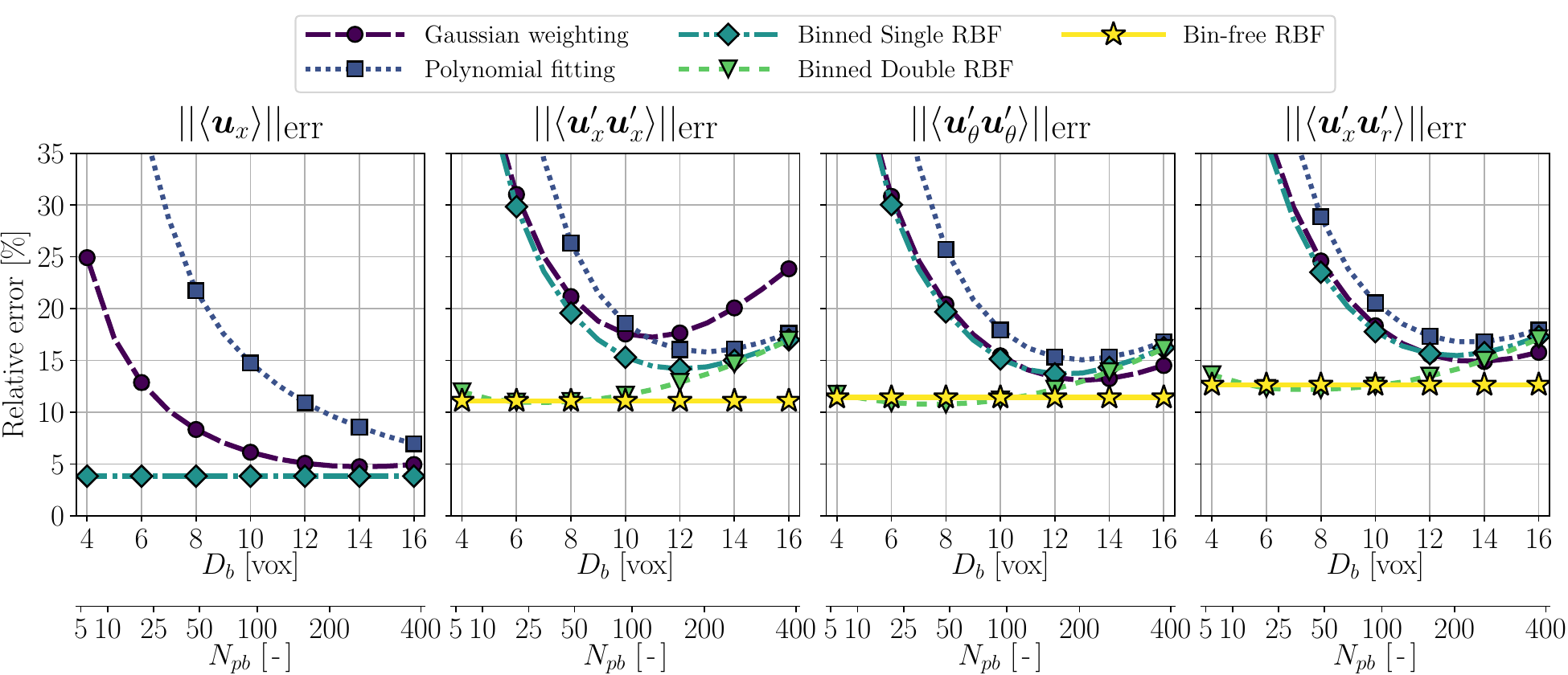}\\
    \hspace{0.5cm}\textbf{(a)} \hspace{3.45cm} \textbf{(b)} \hspace{3.45cm} \textbf{(c)} \hspace{3.45cm} \textbf{(d)} \hspace{-0.05cm}
    \caption{Test case 2. Comparison of the errors for different binning diameters. The five curves correspond to the Gaussian weighting (\protect\markerone, \cite{Aguei1987}), local polynomial fitting (\protect\markertwo \cite{Agueera2016}), Binned Single RBF (\protect\markerthree), Binned Double RBF (\protect\markerfour), and Bin-free RBF (\protect\markerfive). The error norms are the same as the ones defined in equation \eqref{eq:error_norm} and in the first figure, the Binned Double and Bin-free RBF have the same line as Binned Single RBF}
    \label{fig:case2_convergence}
\end{figure*}

The purpose of this test case was to compare and benchmark the methods discussed in Section \ref{sec:5} on a 3D dataset for which the ground truth is available. {We use \num{121500} pseudo-random Halton points as collocation points
(see \citet{Fasshauer2007} for a discussion on random collocation in meshless RBF methods). This gives approximately $8$ particles per basis, in line with the optimal densities identified in \cite{Sperotto2022a}). The RBFs use a fixed radius of 30\,vox.} The PUM used 175 regularly spaced patches with an overlap of $\delta = 0.25$, and no physical constraints were imposed.
All methods with binning use spherical bins of different diameters $D_b$, also spaced on a regular grid. The Reynolds stress regression has the same RBF and patch placement as the mean flow regression. The RBF processing parameters are summarized in Table \ref{tab:case2_rbf_parameters}. All the bin-based approaches use the same binning with size $D_b$ while the Gaussian weighting has a size of $\sigma = D_b / 3$. All five methods are compared on the binning grid.

\begin{table}[t]
    \centering
    \caption{RBF parameters of all three methods for the 3D synthetic jet with Reynolds stress computation}
    \vspace{0.1cm}
    \begin{tabularx}{\columnwidth}{@{}p{0.6\columnwidth}p{0.35\columnwidth}@{}}
    \hline \\[-7pt]
    \textbf{Regression of mean flow} & \\
    \hline
    Number of training points $n_*$ & $1 \cdot 10^6$ (scattered)\\
    Binning diameter $D_b$ & \numrange{4}{16} vox \\
    Number of points per bin $N_{p,b}$ & \numrange{7}{477} \\
    Number of RBFs $n_b$ & \num{121500} (Halton) \\
    RBF radius & 30\;vox\\
    Condition number $\kappa_L$ & $10^{12}$ \\
    Number of patches $M$ $(x \times y \times z)$ & 175 ($7 \times 5 \times 5$) \\
    Overlap $\delta$ & 0.25 \\
    Noise level $q$ & Uniform, $\SI{10}{\percent}$ \\
    \hline \\ [-8pt]
    \textbf{Regression of Reynolds stresses} & \\
    \hline
    Number of training points $n_*$ & \\
    \hspace{0.1cm} Binned Double RBF $(x \times y \times z)$ & \num{288000} ($80 \times 60 \times 60$) \\
    \hspace{0.1cm} Bin-free RBF & $1 \cdot 10^6$ (scattered) \\
    Binning diameter $D_b$ & \numrange{4}{16} vox \\
    Number of points per bin $N_{p,b}$ & \numrange{7}{477} \\
    RBF placement & Regular \\ 
    Number of RBFs $n_b$ & \num{121500} (Halton) \\
    RBF radius & 30\;vox\\
    Condition number $\kappa_L$ & $10^{12}$ \\
    Number of patches $M$ $(x \times y \times z)$ & 175 ($7 \times 5 \times 5$) \\
    Overlap $\delta$ & 0.25 \\
    [3pt]
    \hline
    \end{tabularx}
    \label{tab:case2_rbf_parameters}
\end{table}

Figure \ref{fig:case2_convergence} shows the errors for different statistics defined as:
\begin{align}
    \vert\vert \bm{u} \vert\vert_\text{err} = \frac{\vert\vert \langle \widetilde{\bm{u}} \rangle - \bm{u}_{gt} \vert\vert}{\vert\vert \bm{u}_{gt} \vert\vert}\,,
\end{align}where $\bm{u}$ is either a mean or Reynolds stress and $u_{gt}$ the corresponding ground truth.

Figure \ref{fig:case2_convergence}(a) shows the resulting errors over the binning diameter $D_b$ of the axial mean flow $\langle \bm{u}_x \rangle (\bm{X}_\text{bin})$. The abscissa shows the bin diameter and the average number of particles $N_{pb}$ in each bin. All three RBF-based methods use the same, single regression for the mean which is why they are displayed as one curve. The curve is constant since the regression of the ensemble does not use any binning. For small bin sizes, the error of the Gaussian weighting and polynomial fitting quickly exceeds \SI{15}{\percent}, although the former has a consistently smaller error. This is because of the small number of points within the bin which are insufficient for averaging and local fitting. At the maximum bin size of 16\,vox, the Gaussian weighting reaches an error comparable to the error of the RBF regression whereas the polynomial fitting reaches a minimum of only \SI{8}{\percent}. This is due to the small gradients in the mean flow; for stronger gradients, the spatial low-pass filtering due to larger bin sizes leads to increased error.

\begin{figure*}[t]
    \center
    \includegraphics[width=0.99\textwidth]{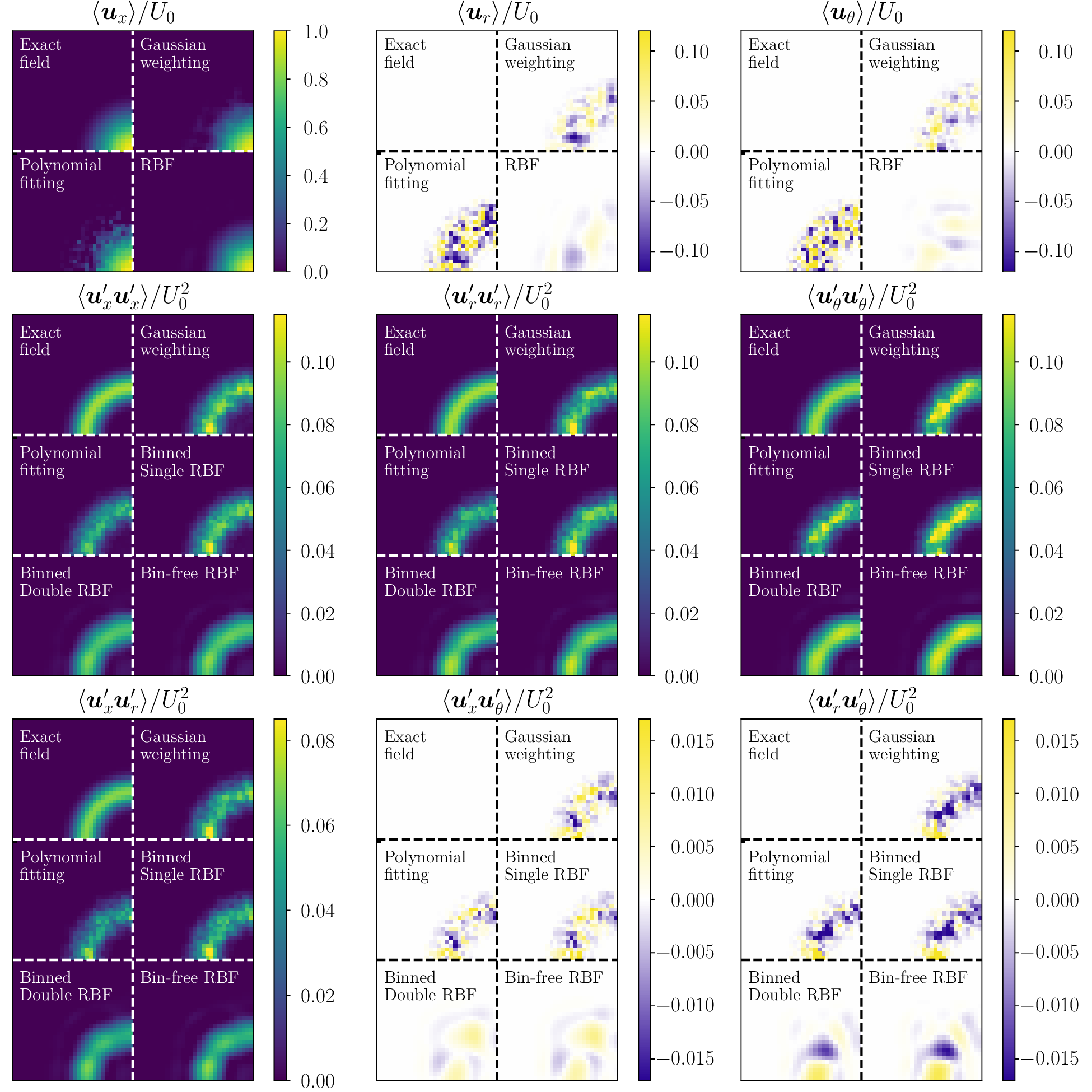}
    \caption{Test case 2. Resulting fields of the slice at $x = 0\,$vox for $D_b = 10\;\text{vox}$. Top row: Mean velocity fields. Middle row: Normal stress fields. Bottom row: Shear stress fields. The four panels for the mean fields show from top left to bottom right: The analytical solution, the Gaussian weighting (\cite{Aguei1987}$, \sigma = D_b / 0.33)$, local polynomial fitting \citep{Agueera2016} and the solution of the RBF regression. For the Reynolds stresses, there are three panels for the RBFs corresponding to the three different algorithms outlined in Subsection \ref{sec:5p2}}
    \label{fig:case2_fields}
\end{figure*}

For the Reynolds stresses, the low-pass filtering due to the binning is more evident. The errors on the stresses $\langle \bm{u}_x^{\prime} \bm{u}_x^{\prime} \rangle (\bm{X}_\text{bin})$, $ \langle \bm{u}_\theta^{\prime} \bm{u}_\theta^{\prime} \rangle (\bm{X}_\text{bin})$, and $ \langle \bm{u}_x^{\prime} \bm{u}_r^{\prime}(\bm{X}_\text{bin}) \rangle$ are shown in subfigures (b)-(d). For the axial Reynolds stress, the spatial inhomogeneities lead to increased errors for larger bins for all methods except the bin-free RBFs. For the axial normal stress, the effects of unresolved mean flow gradients become apparent as the error of the Gaussian weighting strongly increases for large bin sizes. For the other stresses, the mean flow gradients are not as impactful and the weighting mitigates the spatial inhomogeneities. The error trends for the polynomial fitting, Binned Single and Binned Double RBF collapse for bin sized above \SI{12}{\voxel}, because there are no convergence problems and the method of mean subtraction has little influence. However, for bin sizes below \SI{10}{\voxel}, the error quickly reaches values above \SI{15}{\percent} because of poor statistical convergence. For the Binned Double RBF, the unconverged Reynolds stresses are smoothed, preserving the error between 11 and \SI{13}{\percent} for all bin sizes between $D_b =$\;\qtyrange{4}{10}{\voxel}.

The Bin-free RBF outperforms all methods with a constant error of \SI{11}{\percent}, which is the best error achieved by the Binned Double RBF. The fact that the Binned Double RBF converges to the Bin-free RBF at small binning diameters is not surprising, considering that both approximate local statistics. For small diameters (around \SI{8}{\voxel}), the binning only produces a poor approximation of the local statistics and the subsequent regression yields a strong improvement.  

A very similar trend is visible in the tangential Reynolds stress in the third column of Figure \ref{fig:case2_convergence}. The errors for the polynomial fitting and all RBF-based methods appear almost identical to the axial stress. In comparison, the Gaussian weighting reaches its smallest value for the largest bin size. This is because there is no unresolved mean flow gradient which affects the Gaussian weighting. The weight again mitigates spatial inhomogeneities but the error is still \SI{2}{\percent} larger than the smallest value of Binned Double and Bin-free RBF. The correlation between the radial and axial component $\bm{u}_x^\prime \bm{u}_r^\prime$ in the fourth column has the same trend as the tangential Reynolds stress. However, all errors are slightly increased by about \SI{2}{\percent} w.r.t. the other two stresses. The exact reason for this is not known. Yet, the correlation is equally well recovered by all five methods and none of them show additional advantages in this case.

For $D_b = 10\;\text{vox}$, Figure \ref{fig:case2_fields} shows a slice through the jet at $x = 2.5\;\text{vox}$. The first, second and third row contain the mean flow, normal stress and shear stress, respectively. The subfigures of the mean flow contain four panels which are from top left to bottom right: The field of the analytical solution, the Gaussian weighting \citep{Aguei1987}, the local polynomial fitting \cite{Agueera2016} and the mean from the RBF regression. The results of all three mean flow components are similar between all methods. The shape of the mean profile is recovered well and the fields appear slightly noisy in the regions of high shear. As expected from the error curves in Figure \ref{fig:case2_convergence}, the polynomial fitting and Gaussian weighting appear more noisy than the RBF regression. Furthermore, the spikes of the former two methods are random, whereas the RBF regression yields a globally smooth expression.

\begin{table}[t]
    \centering
    \caption{RBF parameters of all three methods for the 3D experimental jet with Reynolds stress computation}
    \vspace{0.1cm}
    \begin{tabularx}{\columnwidth}{@{}p{0.56\columnwidth}p{0.41\columnwidth}@{}}
    \hline \\[-7pt]
    \textbf{Regression of mean flow} & \\
    \hline
    Number of training points $n_p$ & $3.35 \cdot 10^6$ (scattered) \\
    Number of RBFs $n_b$ &  \num{77175} (Halton)\\
    RBF radius & $0.5D$ ($110$\;vox)\\
    Number of solenoidal constraints & \num{14130} ($81 \times 53 \times 23$) \\
    \hspace{0.1cm} $n_\nabla$ $(x \times y \times z)$, outer hull &  \\
    Divergence penalty $\alpha_\nabla$ & 1 \\
    Condition number $\kappa_L$ & $10^{12}$ \\
    Number of patches $M$ $(x \times y \times z)$ & 1300 ($\numproduct{20 x 13 x 5}$) \\
    Overlap $\delta$ & 0.25 \\
    \hline \\ [-8pt]
    \textbf{Regression of Reynolds stresses} & \\
    \hline
    Number of training points $n_p$ & \\
    \hspace{0.1cm} Binned Double RBF & \num{760725} ($\numproduct{161 x 105 x 45}$) \\
    \hspace{0.1cm} Bin-free RBF & $3.35 \cdot 10^6$ (scattered) \\
    Number of RBFs $n_b$ &  \num{77175} (Halton)\\
    RBF radius & $0.5D$ ($110$\;vox)\\
    Binning diameter $D_b$ & $0.15D$ (44\;vox)\\
    Condition number $\kappa_L$ & $10^{12}$ \\
    Number of patches $M$ $(x \times y \times z)$ & 1300 ($\numproduct{20 x 13 x 5}$) \\
    Overlap $\delta$ & 0.25 \\
    [3pt]
    \hline
    \end{tabularx}
    \label{tab:expjet_parameters}
\end{table}  

The subfigures of the Reynolds stresses additionally contain two panels at the bottom showing the result of the Binned Double and Bin-free RBF method. The effects of not subtracting the local mean are evident in the second row, which shows the three normal stresses. In the core of the jet, in the bottom right region of the panel, the Gaussian weighting has a non-zero axial normal stress $ \langle \bm{u}_x^{\prime} \bm{u}_x^{\prime} \rangle$ in regions where it should be zero. We attribute this to mean flow gradients within the bin. The other four methods are not affected by this. Moreover, we again highlight the smoothing properties of the second RBF regression observed in the contours of the normal stresses obtained by the Binned Double and Bin-free RBF. 

The same observations hold for the shear stresses in the bottom row of the figure. All methods recover the correlation well. The Gaussian weighting is most severely affected by convergence issues whereas the top-hat approach, polynomial fitting and Binned Single RBF have almost the same shear stress fields. 

To conclude this section, the methods based on two successive RBF regressions perform the best for the analyzed test case. For small binning diameters, Binned Double RBF and Bin-free RBF yield almost the same result as the binning only introduces a slight modulation. Besides the lowest error, the RBF regressions also give continuous expressions of the statistics which enables super-resolution and analytical gradients for all Reynolds stresses.

\begin{figure*}[t]
    \center    \includegraphics[width=0.99\textwidth]{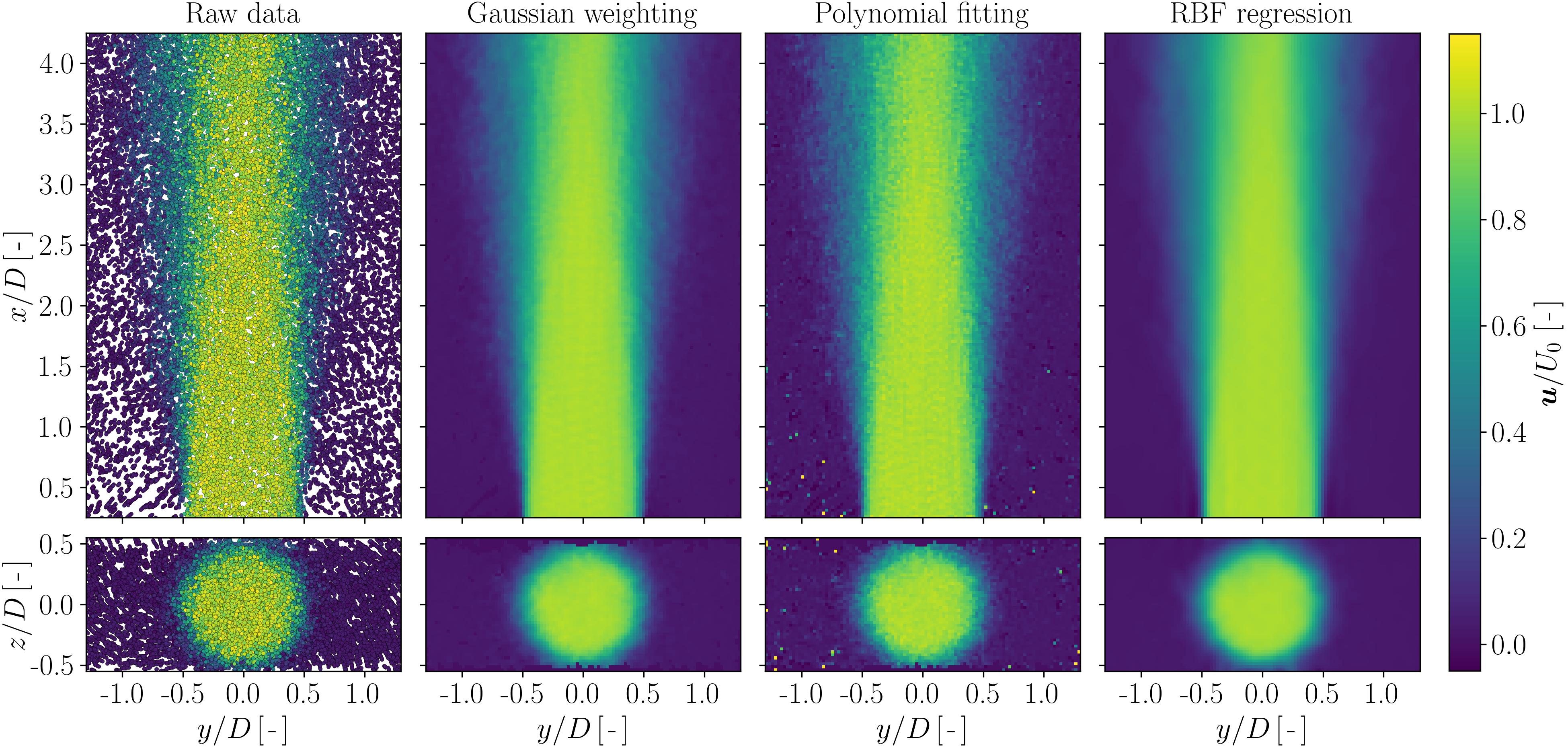}\\
    \hspace{0.1cm}\textbf{(a)} \hspace{3.11cm} \textbf{(b)} \hspace{3.11cm} \textbf{(c)} \hspace{3.11cm} \textbf{(d)} \hspace{0.6cm}
    \caption{Test case 3. Resulting fields of the axial flow component in a slice through the regression volume. Top row: vertical slice at $z/D = 0$. Bottom row: horizontal slice at $x/D = 2$. Subfigure \textbf{(a)} shows the scattered training data $\bm{u}_*$ in a thin volume around the slice and subfigures \textbf{(b)}-\textbf{(d)} respectively show the mean field $\langle \bm{u} \rangle / U_0$ from: the Gaussian weighting (\cite{Aguei1987}$, \sigma = D_b / 0.33)$, local polynomial fitting \citep{Agueera2016} and the RBF regression of the ensemble}
    \label{fig:case3_fields_mean}
\end{figure*}

\subsection{3D Experimental turbulent jet}
\label{sec:jet_results}

\begin{figure*}[t]
    \center
    \includegraphics[width=0.99\textwidth]{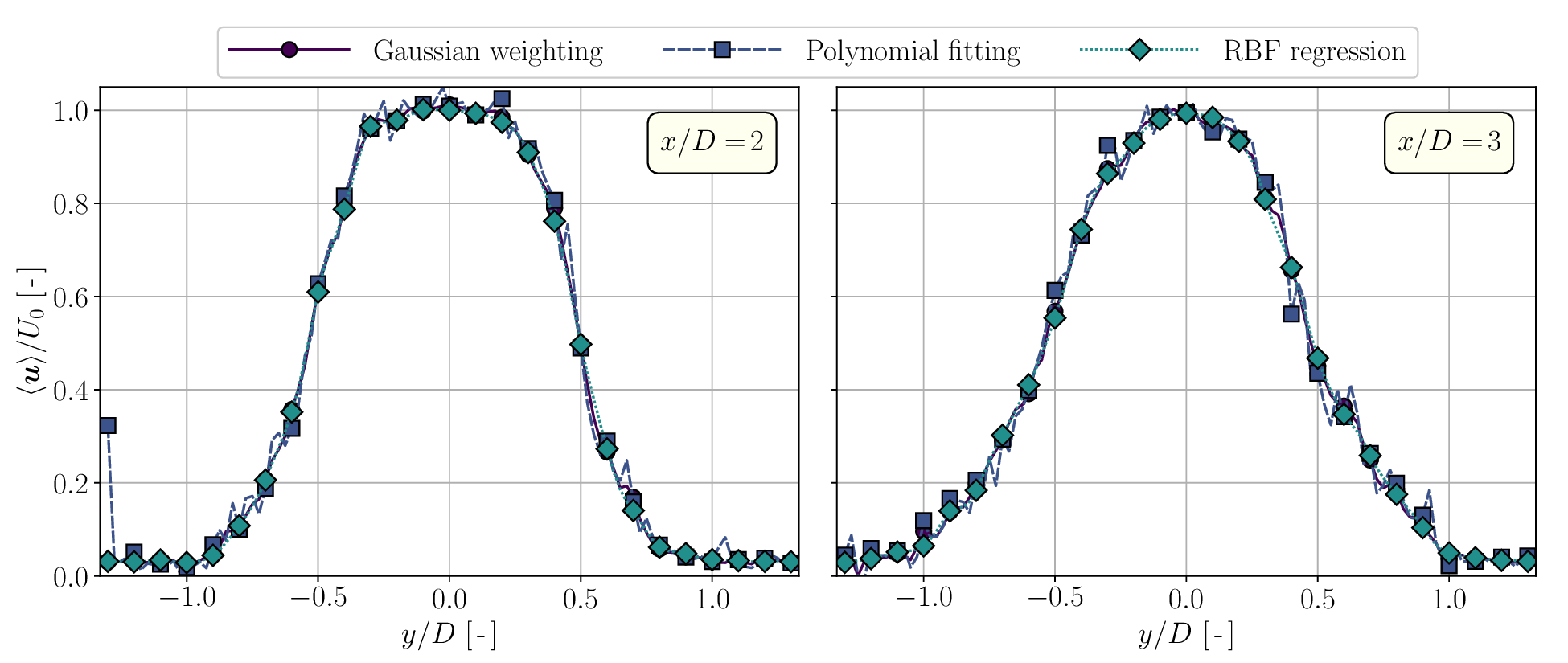}
    \caption{Test case 3. Resulting mean velocity profile $\langle \bm{u} \rangle / U_0$ extracted at $z/D = 0$ and $x/D = 2$ (left) and $x/D = 3$ (right). The three curves correspond to the Gaussian weighting (\protect\markerone, \cite{Aguei1987}), local polynomial fitting (\protect\markertwo, \cite{Agueera2016}) and the RBF regression of the ensemble (\protect\markerthree)}
    \label{fig:case3_profiles_mean}
\end{figure*}


\begin{figure*}[t]
    \center
    \includegraphics[width=0.99\textwidth]{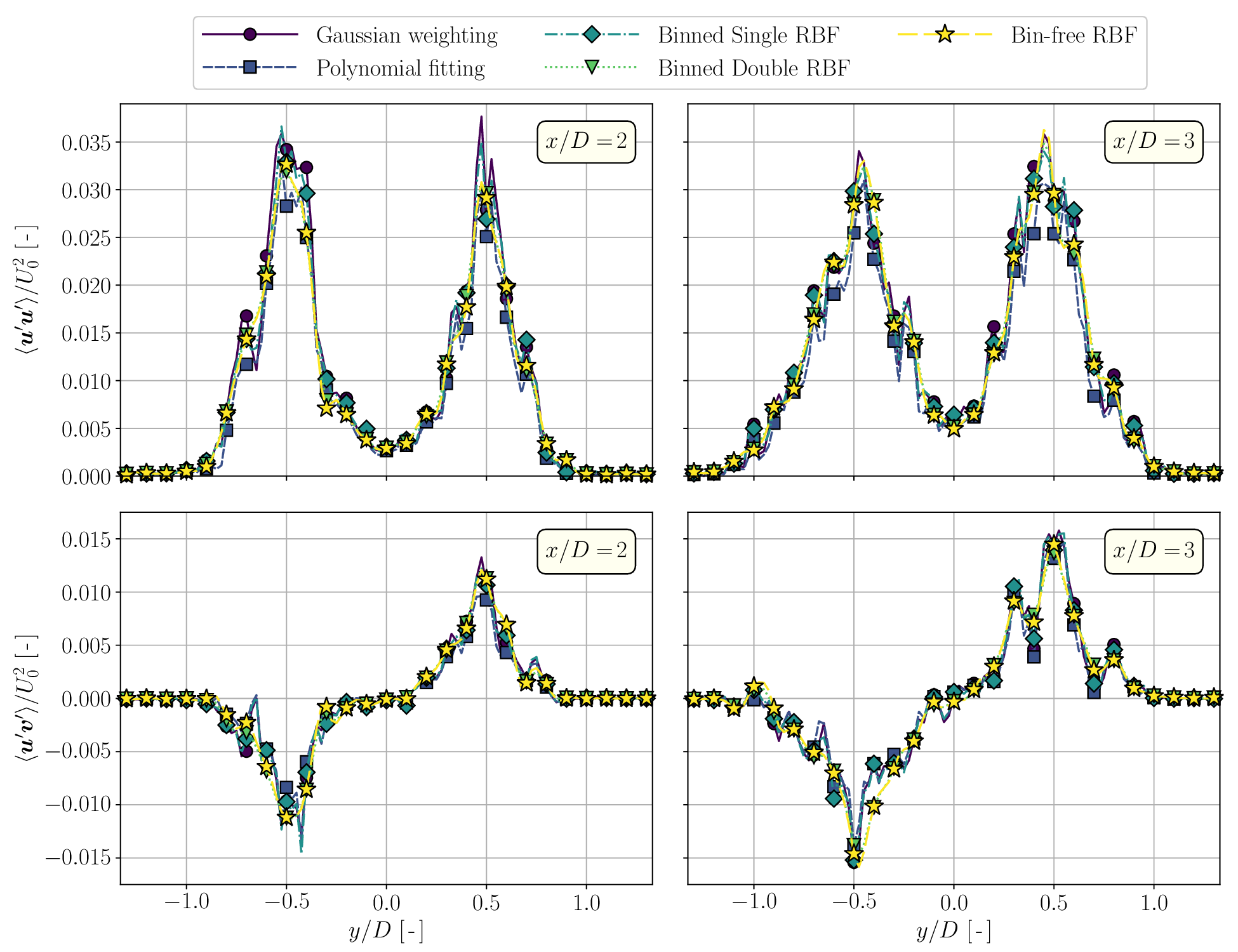}
    \caption{Test case 3. Resulting Reynolds stress profile $\langle \bm{u}^\prime \bm{u}^\prime \rangle / U_0^2$ (top) and $\langle \bm{u}^\prime \bm{v}^\prime \rangle$ (bottom) extracted at $z/D = 0$ and $x/D = 2$ (left) and $x/D = 3$ (right). The five curves correspond to the Gaussian weighting (\protect\markerone, \cite{Aguei1987}), local polynomial fitting (\protect\markertwo, \cite{Agueera2016}), Binned Single RBF (\protect\markerthree), Binned Double RBF (\protect\markerfour), and Bin-free RBF (\protect\markerfive)}
    \label{fig:case3_profiles_reynolds_stresses}
\end{figure*}

{The regression of the mean flow field was done with $n_b=\num{77175}$ RBFs, placed with pseudo-random Halton points as in the previous test case. Considering the measurement volume of $V\approx\SI{4000}{\milli\meter^3}$, this yields an RBF density of $\rho_b=n_b/V\approx 1.9$ bases per $\mbox{mm}^3$. For a uniform distribution of points, using geometric probability one could thus estimate an expected average distance of $\mathbb{E}=(4/3)^{1/3} \rho_b^{-1/3}\approx \SI{0.9}{\milli\meter}$ between bases, enabling sufficient overlapping if these have a radius of $0.5D$.} Divergence-free constraints were imposed in \num{14130} points on the outer hull of the measurement domain, and a penalty of $\alpha_\nabla = 1$ was applied in the whole flow domain. {The imposed constraints do not significantly impact the $l_2$ norm of the error, but allows for better derivatives and improve the computation of derived quantities such as pressure \cite{Sperotto2022a}.} In total, 1300 patches were used for the PUM, again with an overlap of $\delta = 0.25$. For the computation of the Reynolds stresses, \num{760725} bins with a diameter of $0.15D$ were placed on a regular grid of $\numproduct{161 x 105 x 45}$ points in $x \times y \times z$. This yielded an average of 65 vectors within each bin. 
The second regression reused the same basic RBF and PUM settings. All processing parameters are summarized in Table \ref{tab:expjet_parameters}.

Figure \ref{fig:case3_fields_mean} shows slices of the velocity field from the PTV data $\bm{u} / U_0$ and the computed mean $\langle \bm{u} \rangle / U_0$ for each algorithm. The slices are respectively taken from two planes at $z/D = 0$ and $x/D = 2$. The raw data in a thin volume around the slice is shown as a scatter plot in subfigure (a) while subfigures (b)-(d) show the velocity on the binning grid. All three methods capture the spreading of the symmetric jet well although the RBF regression appears smoother, particularly in the shear layer. The horizontal slice at $z/D = 0$ further confirms this lack of convergence as the bins on the domain boundary are particularly noisy. In contrast, the RBF solution shows a smooth behaviour, as the divergence-free flow acts as a regularization which prevents sharp, noisy spikes.

Figure \ref{fig:case3_profiles_mean} shows two mean velocity profiles, extracted at $z/D = 0,\,x/D = 2$ (left) and $z/D = 0,\,x/D = 3$ (right). It can be very well seen that the profiles for all three methods almost collapse. The profiles are not symmetric around around the central axis but this asymmetry is equal between all methods, so we attribute it to the jet facility and not the methods. The RBF method yields the best performance in the aforementioned regions of low particle seeding. While the other two methods produce spikes in the mean flow due to problems in the statistical convergence, the RBFs yield a smooth profile of the axial mean velocity.

The Reynolds stress profiles in Figure \ref{fig:case3_profiles_reynolds_stresses} show the same characteristics as the mean flow. We show the normal stress $\langle \bm{u}^{\prime} \bm{u}^{\prime} \rangle / U_0^2$ and the shear stress $\langle \bm{u}^\prime \bm{v}^\prime \rangle/ U_0^2$ in the top and bottom row, respectively. All methods give results which agree with theoretical expectations: the stresses are largest in the shear layer and expanding with the jet. Furthermore, the normal stress is an even function while the shear stress is an odd function. Yet, the Reynolds stresses appear more noisy than the mean flow, as convergence is slower for higher order statistics. This is particularly visible in the Gaussian weighting approach, which shows significant spikes in each of the four subfigures with differences in the peak amplitude compared to the other methods. In contrast, polynomial fitting and Binned Single RBF have almost the same curve. The lack of convergence is mainly responsible for the non-smooth profile rather than the specific method of mean subtraction. The Binned Double RBF and Bin-free RBF yield smoother curves compared to the other methods but still struggle in specific areas, like $(x/D, y/D) = (3, 0.35)$ for $\langle \bm{u}^\prime \bm{v}^\prime \rangle$ where the profiles have an unexpected kink. Yet, this kink is also visible for all other methods and likely stems from unfiltered outliers or a general lack of points in this region.

To conclude, the two successive RBF regressions give the best results also for the experimental test case. In regions with sparse or noisy data, the regularization yields a smooth solution and matches the binning-based approaches in all other regions.

\section{Conclusions and Perspectives}\label{sec:7}

We propose a meshless and binless method to compute statistics in turbulent flows in ensemble particle tracking velocimetry (EPTV). We use radial basis functions (RBFs) to obtain a continuous expression for first and second-order moments. We showed through simple derivations that an RBF regression of a statistical field is equivalent to performing spatial averaging in bins. We expanded this idea and showed averaging the weights from multiple regressions can be approximated with a single, large regression of the ensemble of points. The test case of a 1D Gaussian process served as numerical evidence to prove the convergence of the weights and the solution. The resulting matrix is very large, and the computational cost of inverting is prohibitive. Therefore, we employ the partition of unity method (PUM) and the RBFs to reduce the computational cost significantly. Together, both approaches result in analytical statistics at a low cost, even for large-scale problems.

We proposed three different RBF-based approaches and compared them with existing methods, namely Gaussian weighting \cite{Aguei1987} and local polynomial fitting \cite{Agueera2016}. The proposed methods range from simple ideas based on existing literature \citep{Agueera2016} to a fully mesh- and bin-free method which uses two successive RBF regressions. On a synthetic test case, the RBF-based methods outperformed the methods from existing literature in both first- and second-order statistics, with the bin-free method having the lowest error. Therefore, besides giving an analytical expression, the bin-free methods also require less data for convergence, which is highly relevant for experimental campaigns.

The same conclusions hold on experimental data, with the RBF approaches producing the best results. All methods show a qualitative agreement with literature expectations with the binning-based approaches having more noise. Insufficient convergence within a bin results in spikes, whereas the methods with a second regression yield a smooth curve with almost no outliers. Therefore, the two successive regressions have the double merit of providing smooth and noise-free analytical regression that can be used for super-resolution of the flow statistics.

Ongoing work focuses on integrating the pressure Poisson equation in the Reynolds Averaged Navier Stokes framework to obtain the mean pressure field. This can be done with a mesh-free integration following the initial velocity regression, or by coupling both steps in a non-linear method.


\begin{acknowledgements}
The authors kindly acknowledge Alessia Simonini for her help in setting up and conducting the experiments. The authors also acknowledge David Hess from Dantec Dynamics for his support in setting up the measurement system and helping during the post-processing with Dynamic Studio.
\end{acknowledgements}

\begin{funding}
The authors gratefully acknowledge the financial support of the ‘Fonds de la Recherche Scientifique (F.R.S. -FNRS)’ for the FRIA grant with grant
number FC57471 supporting the PhD of Mr. Ratz.
\end{funding}





\bibliographystyle{spbasic}
\bibliography{Ratz_and_Mendez_2024.bib}

\appendix




\end{document}